\documentstyle[11pt,aaspp4]{article}

\def\plotfiddle#1#2#3#4#5#6#7{\centering \leavevmode
    \vbox to#2{\rule{0pt}{#2}}
    \includegraphics{#1}}

\def\etal{{\it et~al.\ }}

\begin{document}

\title{The Distribution of High Redshift Galaxy Colors: Line of Sight
Variations in Neutral Hydrogen Absorption}

\author{Matthew A. Bershady}

\affil{Department of Astronomy, University of Wisconsin \\ 475 North
Charter Street, Madison, WI 53706 (mab@astro.wisc.edu)}

\author{Jane C. Charlton}

\affil{Department of Astronomy \& Astrophysics and \\ Center for
Gravitational Physics \& Geometry, Pennsylvania State University, \\
University Park, PA 16802 (charlton@astro.psu.edu)}

\author{Janet M. Geoffroy}
\affil{Department of Astronomy \& Astrophysics, Pennsylvania State
University \\ University Park, PA 16802 (jmg@astro.psu.edu)}

\begin{center}
{\large\bf Abstract}
\end{center}

We model, via Monte Carlo simulations, the distribution of observed
$U-B$, $B-V$, $V-I$ galaxy colors in the range $1.75<z<5$ caused by
variations in the line-of-sight opacity due to neutral hydrogen
(HI). We also include HI internal to the source galaxies. Even without
internal HI absorption, comparison of the distribution of simulated
colors to the analytic approximations of Madau (1995) and Madau \etal
(1996) reveals systematically different mean colors and scatter.
Differences arise in part because we use more realistic distributions
of column densities and Doppler parameters. However, there are also
mathematical problems of applying mean and standard deviation
opacities, and such application yields unphysical results. These
problems are corrected using our Monte Carlo approach. Including HI
absorption internal to the galaxies generaly diminishes the scatter in
the observed colors at a given redshift, but for redshifts of interest
this diminution only occurs in the colors using the bluest
band-pass. Internal column densities $< 10^{17}$ cm$^2$ do not effect
the observed colors, while column densities $> 10^{18}$ cm$^2$ yield a
limiting distribution of high redshift galaxy colors. As one
application of our analysis, we consider the sample completeness as a
function of redshift for a single spectral energy distribution (SED)
given the multi-color selection boundaries for the Hubble Deep Field
proposed by Madau \etal (1996). We argue that the only correct
procedure for estimating the $z>3$ galaxy luminosity function from
color-selected samples is to measure the (observed) distribution of
redshifts {\it and} intrinsic SED types, and then consider the
variation in color for each SED and redshift. A similar argument
applies to the estimation of the luminosity function of
color-selected, high redshift QSOs.


\section{Introduction}

It has been roughly a quarter of a century since the ultraviolet
opacity of the Universe caused by neutral hydrogen first was used to
constrain the number of high redshift objects (Partridge 1974, Davis
\& Wilkinson 1974, Koo \& Kron 1980). These and other early searches,
summarized by Koo (1986), were at red observed wavelengths, and aimed
at probing redshifts generally in excess of five. With the advent of
deep $U$ band imaging over relatively large areas (e.g. Koo 1981),
over the last decade a broad-band technique has been considered at
shorter observed wavelengths for constraining the galaxy population
above somewhat lower redshifts (of order 3, Majewski 1988,
Guhathakurta \etal 1990). Specifically, objects were sought which had
excessive diminution of their $U$ band flux relative to redder
bands. This has been referred to as the $U$-band ``drop-out''
technique, although the ``drop-out'' technique can be applied to
redder bands, depending on the redshift range of interest.

However, it has been only recently that Steidel and collaborators have
brought this technique to fruition: with the advent of large-format
blue-sensitive CCDs and large (10m) ground-based telescopes, it has
been possible for this group to select $U$-band drop-outs at the
unprecendented depths of $R = 25$, {\it and} to spectroscopically
confirm their redshifts at $z>2.7$ (e.g. Steidel \etal 1996, Lowenthal
\etal 1997). This has raised the real possibility that the comoving
properties of the galaxy population can be studied at extremely early
times in great detail (Steidel \etal 1998, Dickinson 1998).

The current method of selecting high redshift galaxies is a rather
simple one which relies on defining regions in one or more two-color
diagrams corresponding to the expected range of high redshift galaxy
colors (e.g. Steidel \etal 1993, 1995). Following along lines similar
to what was done over twenty years ago (e.g. Meier 1976), these
regions are defined by simulating the expected colors of high redshift
galaxies based on evolutionary spectral synthesis models (currently,
for example, Bruzual \& Charlot 1993) and the mean line of sight (LOS)
opacity due to neutral hydrogen (Madau, 1995). Madau \etal (1996) have
refined this technique for the specific broad passbands used with Wide
Field Planetary Camera (WFPC2) to observe the Hubble Deep Field (HDF),
and for a variety of galaxy spectral energy distributions (SEDs). For
QSOs, Giallongo \& Trevese (1990) have performed a similarly thorough
analysis in the context in of ground-based photographic photometry.

Fortuitously, and somewhat by design, the regions of color space
searched for high redshift galaxies largely avoid the regions of color
space inhabited by Galactic stars and galaxies at lower redshifts.
Consequently, the current technique has been very efficient
(i.e. reliable) for finding high redshift galaxies. As such, the
selection method is sufficient, e.g. for studying large scale
structure at high redshift (Steidel \etal 1998). The details of the
selection procedure are critical, though, for understanding the true
comoving number and distribution of galaxy types at these high
redshifts. Comparably little attention has been paid, for example, to the
completeness of the current samples. Indeed, Lowenthal \etal (1997)
found a number of high redshift galaxies selected nearby, but not
within, the nominal region of high-redshift color space within the
HDF. On an empirical basis alone, this indicates that the
current selection methods may be somewhat incomplete.

There are several other reasons to expect that the precise selection
function for high redshift galaxies is complicated and not yet well
defined. While variations in SEDs play a critical role in determining
the color-based selection criteria for selecting high redshift
galaxies, the intervening opacity is as important. To meet the
``drop-out'' criterion in a two-color diagram, ideally the bluest band
(alone) samples below 1216 \AA \ in the galaxy's rest-frame. In this
situation, variations in galaxy age, metallicity and dust (reddening)
will cause the largest changes in the redder color due to changes in
the slope of the ultraviolet continuum. While in principle the
selection criterion can be tailored for a specific intrinsic spectral
type, the effects of obscuration by dust in intervening systems must
also be considered (e.g. Heisler and Ostriker 1988; Fall and Pei 1993
-- in the context of QSOs). The bluer color, however, will be
dominated by the effect of the LOS opacity due to neutral hydrogen.
What is worrisome here is that the dominant contribution to the
continuum opacity is from a small number of absorbers at the largest
column densities in a given LOS (Madau \etal 1996). This raises the
likely possibility that small number statistics will cause large
variations in the observed colors of high redshift galaxies of the
{\it same intrinsic spectral type}.

While Madau (1995) and Madau \etal (1996) have considered the effects
of LOS variations on the observed high redshift galaxy colors, they
have not done so within the context of a Monte Carlo simulation
including a realistic distribution of absorbers at all column
densities. As we shall demonstrate, there are a number of reasons why
Monte Carlo simulations are necessary for predicting the correct
distribution of observed galaxy colors. Moreover, there are
substantial uncertainties in the known distribution of column
densities and Doppler $b$ parameters of intervening absorbers. Recent
results from the Keck telescope (Kim \etal 1997) provide new estimates
for absorber properties that are an improvement over what has been
adopted in previous work. The spirit of the current analysis is to
look in detail at variations in the expected colors of high redshift
galaxies due to a range of (and variations in) LOS absorption.

For clarity and simplicity, we have adopted a single, representative
galaxy SED to illustrate the results of our simulations; we avoid
considering the quantitative effects of a range of SEDs on the
observed color distribution, as has Madau \etal (1996). However our
methodology can be be carried over in a general way to a realistic
ensemble of SEDs for any source type (e.g. galaxies or QSOs), as we
will discuss at the end of the paper. In the next section we describe
our Monte Carlo method and the range of adopted absorber parameters,
and compare it to that of Madau's (1995). We compare the results of
these two methods in \S 3 in terms of mean and scatter in colors due
to variations in line-of-sight attenuation. Section 4 illustrates the
full range of colors for a single SED over a range in
redshift. Section 5 contains an application of our method to
estimating the selection completeness of high redshift galaxies.  In
\S 6 we consider the effects of internal neutral hydrogen absorption.
The results of this work, presented in the HDF filter system, are
independent of cosmological parameters (e.g. q$_0$ or H$_0$), and are
summarized in \S 7.


\section{Simulations of Intervening HI Absorption and
  High Redshift Galaxy Colors}

To model the colors of high redshift galaxies, we begin with an input
spectrum, shift it appropriately for redshift, and then attenuate it
to account for intervening neutral hydrogen. Our method of applying
the attenuation is to perform Monte Carlo simulations of many lines of
sight to produce an ensemble of attenuated spectra, for which colors
are calculated independently. Each line-of-sight attenuation
represents a random sample of discrete absorbers selected from
distributions constrained by observations of quasar absorption line
systems.  In the following subsections we describe a computationally
efficient realization of our method and contrast it to the existing
one used by Madua (1995). The specific model input parameters are also
described.

With a red-shifted, attenuated spectrum in hand, we compute colors by
convolving the spectrum with the Wide Field Planetary Camera 2 filter
responses F300W, F450W, F606W, and F814W (averaged over the four
detectors). We refer to these hereafter as the $U_{300}$, $B_{450}$,
$V_{606}$, and $I_{814}$ bands. The response curves include the filter
transmission as well as estimates of the total system throughput
(Rudloff \& Baggett 1995); the known red leak in the F300W filter is
included. They are illustrated in Figure 1, along with a
representative galaxy spectrum at $z=3$ (attenuated and unattenuated),
which is described below in \S2.2 and \S2.3.2.  Colors are calculated
using zeropoints from the AB system; in this system an A0V star has
colors of $(U_{300}-B_{450}) = 1.42$, $(B_{450}-V_{606}) = -0.20$, and
$(V_{606}-I_{814}) = -0.31$.


\subsection{The Mean Line-of-Sight Method}

The only method previously used for calculating the mean colors of
high redshift galaxies has been to convolve the mean attenuation
curves presented, e.g., by Madau (1995) with a chosen SED and
calculate the colors of the resulting mean spectrum. A mean
attenuation curve is generated by integrating over the probability
distribution of absorbers over all lines of sight. The model
parameters include redshift, and the distribution functions of column
density as a function of redshift and Doppler $b$ parameter, although
in Madau's formulation the $b$ parameter was held constant at 35
km s$^{-1}$. The mean attenuation curve can be computed from (Madau
1995):

\begin{equation}
\left< e^{-\tau} \right> = \int e^{-\tau} p(\tau) d\tau =
  \exp \left[ \int _{0} ^{z_{em}} \int \frac{\partial^2 N}{\partial
  N_{HI}
\partial z} \left( 1-e^{-\tau_c} \right) dN_{HI} dz \right],
\end{equation}

\noindent where $p(\tau)$ is the probability distribution of total
optical depths, $z_{em}$ is the emission source redshift, and $N_{HI}$
is the column density of neutral Hydrogen. The specific input
parameters used by Madau are described below. Similarly, the method
proposed to estimate the range of colors due to line-of-sight
variations is to use the line-of-sight variance in attenuation in a
similar fashion as $\left< e^{-\tau} \right>$:

\begin{equation}
\sigma^2 \left(e^{-\tau} \right) = \int \left( e^{-\tau} - \left<e^{-\tau}
\right> \right)^2 p(\tau) d\tau = \exp \left[ - \int_{0}^{z_{em}} \int
\frac{\partial^2 N}{\partial N_{HI} \partial z} \left(1-e^{-2\tau_c}\right)dN_{HI}
dz \right] - e^{-2\tau_{eff}},
\end{equation}

\noindent where $\tau_c$ is the optical depth of an individual cloud,
and $\tau_{eff} = -\ln \left( \left< e^{-\tau} \right> \right)$ is the
mean effective optical depth.  In other words, $\sigma$ is the
expected scatter around $\left< e^{-\tau} \right>$ due to variations
in the number of absorbers along the line of sight.  In Madau's
formulation $\sigma^2 (e^{-\tau})$ is calculated for a given
wavelength interval, $\Delta\lambda$.  Colors produced with these mean
line-of-sight formulae we refer to as generated from the ``MLOS''
method.

While this method is computationally simple to apply, unfortunately
there are several problems with this procedure.  First, because colors
are logarithms of flux ratios, the colors of the mean attenuated
(galaxy) SED are not the same as the mean of individual attenuated
galaxy colors. That is, the mean of the color distribution for a given
SED cannot be calculated using the mean attenuation curve given by
equation (1).  It is important to realize that the so-called ``mean''
colors calculated using equation (1) will be statistically offset from
the actual means of the color distribution. We quantify this effect in
the next section. A similar argument concerns the application of
equation (2) for deriving the range (standard deviation) of colors.

However, for estimating the range of colors due line-of-sight
variations, there is a more fundamental problem. For broad-band
colors, the relevant variation in attenuation is, to first order, that
within the band-pass, and not in some arbitrary $\Delta\lambda$. (To
second order the distribution of variations in attenuation within the
band-pass is important for determining 'color-terms,' i.e. the fact
that the broad-band filter response curves are not square nor is the
unattenuated spectrum flat in $f_\lambda$.) Indeed, a cursory
examination of Madau's (1995) Figure 3 reveals that the mean
$+1\sigma$ values of $e^{-\tau}$ reach unphysical values of $>1$. At
the $-2 \sigma$ level, this figure also implies that $e^{-\tau} < 0$
at some wavelengths. Part of the problem stems from the fact that
these curves were derived assuming Gaussian statistics.  At any given
wavelength, the number of discrete absorbers making significant
contribution to the attenuation is small; Poisson statistics are
appropriate.  More importantly, the variation in the attenuation in
the MLOS method is calculated within a $\Delta\lambda$ smaller or
comparable to the typical line-width of the absorber. That is, because
the absorbers form a discrete distribution in redshift on any single
line of sight, variations in attenuation over small ranges in
wavelength can be very large. Averaged over large $\Delta\lambda$,
e.g. a broad-bandpass, the variations should be less, as noted by
Madau \etal (1996). However, the correct formulation for the variation
in broad bands has not been done. Finally, as noted above, it is not
valid mathematically to use a single, one--sigma attenuation curve
calculated by the MLOS method to calculate the standard deviation of
the observed colors.

\subsection{Monte Carlo numerical method}

The only correct way to calculate the mean colors of high redshift
galaxies is to calculate {\it the mean of the distribution of
attenuated colors.}  To do this, we create a population of discrete,
neutral hydrogen absorbers and place them along random lines of sight.
These absorbers attenuate the intrinsic (galaxy) SED via Lyman series
lines and the Lyman limit break. Our Monte Carlo (hereafter, MC) approach
simulates an ensemble of attenuated galaxy spectra for a given
redshift and input spectrum. These attenuated spectra are different by
virtue of the details of each line-of-sight opacity. Specifically, for
each simulation, we draw randomly from distributions in column density
(N$_{HI}$), $b$ parameter, and absorber redshift ($z_a$), thus
yielding a unique line of sight. We consider several distribution
functions to determine the effects of different IGM models on observed
galaxy colors.

We have developed two methods for applying the attenuation to a
spectrum. We refer to these as the 'high resolution' (HIGHRES) and
'low resolution' (LOWRES) methods. As one might guess, the LOWRES
method is computationally much faster, while in principle the HIGHRES
method should be more accurate. These models differ in how the
Lyman-series (i.e. forest) attenuation is applied.  Lyman limit
absorption is calculated identically for both methods, using the
equation

\begin{equation}
f=f_0 \exp{\left(-N_{HI} \left(6.3~\hbox{x}~10^{-18}\right)\left(
\frac{\lambda_{obs}} {911.75 \left(1+z\right)} \right)^3 \right)},
\end{equation}

\noindent where $f$ and $f_0$ are the attenuated and unattenuated flux
per unit wavelength, respectively, $N_{HI}$ is the neutral hydrogen
column density (cm$^2$) and $\lambda_{obs}$ is the observed
wavelength.

The HIGHRES method for adding absorption lines to the intrinsic SED
requires an extremely high resolution spectrum (about 0.01 \AA \
spacing). The spectrum is then attenuated for each line using the
standard equation: $f=f_0 e^{-\tau}.$ The HIGHRES procedure adds
lines modeled with a Voigt profile which gives

\begin{equation}
\tau(\lambda) = {{\pi^{1/2} e^2} \over {m_e c^2}} {{N_{HI} f_{osc} 
\lambda_0}
\over {\Delta \lambda_D}} u(x,y),
\end{equation}

\noindent where $f_{osc}$ is the oscillator strength of the transition, 
$\lambda_0$ is its rest wavelength, and $\lambda_D$ is the Doppler
wavelength shift for a gas of temperature $T$ given by

\begin{equation}
\Delta \lambda_D = {\lambda \over c} \left ( {{2kT} \over {m_H}} 
\right )^{1/2}.
\end{equation}

\noindent The Voigt function $u(x,y)$ is the real part of the function
\begin{equation}
w(z) = e^{-z^2} {\rm erfc}\left( -iz \right) = u(x,y) + i v(x,y),
\end{equation}

\noindent where here $x$ is defined as the wavelength difference from line
center in units of the Doppler shift, $x =
\Delta\lambda/\Delta\lambda_D$. The width of the transition $y$ is
also defined in units of the Doppler shift as

\begin{equation}
y = {{\lambda_0^2 \ \Gamma} \over {4 \pi c}} {{1}\over{\Delta\lambda_D}}
\end{equation}

\noindent where $\Gamma$ is the damping constant.

The LOWRES method requires a much lower resolution spectrum ($\sim$4
\AA \ spacing). The spectrum is attenuated for each line using a
triangular approximation to the line shape via the following, simple
numerical steps: (1) Locate the pixel nearest to the actual central
wavelength of the line. (2) Based on the column density and $b$
parameter of the cloud, identify the appropriate equivalent width in a
look-up table pre-calculated based on the curve of growth. (3)
Calculate the area underneath that region of the spectrum that would
need to be eliminated to account for the equivalent width of the
line. (4) Divide the area in step (3) by two and reduce the flux of
the pixels to the left and right by the necessary amount, limited by
$f_\lambda=0.$ (5) If the required area is not fully subtracted in
step (4), continue reducing the flux of the left and right pixels by
the same procedure.

In order to test the accuracy of LOWRES, we produced identical lines
of sight to an identically redshifted galaxy spectrum and calculated
the colors using both HIGHRES and LOWRES. A representative example for
a galaxy at $z = 3$ is shown in Figure 2, over a small range of
observed wavelength. (For clarity, the input spectrum is flat in
$f_\lambda$. Figure 1 contains the LOWRES spectrum over a broaded
wavelength range for a more realistic input spectrum desribed below in
\S2.3.2.) This comparison showed that LOWRES colors are accurate to better
than $\sim$0.5\%, and yet LOWRES requires 2.5\% as much memory and
substantially less computational time. As a consequence, all further
Monte Carlo simulations use LOWRES.

We performed several other tests of our numerical method. For example,
we determined that clouds with column densities below $5\times10^{13}$
cm$^2$ do not noticeably affect the colors (i.e. below 1\%), for a
variety of input column density distributions. Consequently, we
limited the application of attenuation to clouds with
$>5\times10^{13}$ cm$^2$ for our primary model (MC-Kim, described
below). However, for all other models we include clouds with neutral
hydrogen column densities as low as $2 \times 10^{12}$ cm$^{-2}$ to
more directly compare with Madau's models. The attenuation curves
described by Madau (1995) include 17 Lyman series lines. Here we have
used 25 Lyman series lines; we note this difference amounts to an
insignificant $\sim$0.5\% change in the resulting colors in all cases
we considered.

\subsection{Model input ingredients}

\subsubsection{$N_{HI}$ and $b$}

We consider a set of three possible distributions in column density
and $b$ parameter as a function of redshift, realized either via the
mean line-of-sight method (MLOS) or Monte Carlo method (MC). In total,
six separate sets of models are considered, enumerated below.  The
first four are based on versions of the column density distributions
used by Madau (1995) and Madau \etal 1996, that draw on the
observational constraints of Murdoch \etal (1986), Tytler (1987) and
Sargent \etal (1989). These represented the best observational
constraints at the time of Madau's original paper.  More recent
observations of Lu \etal (1996) and Kim \etal (1997) indicate that
substantial revisions in the previously adopted distributions are in
order. In particular, these more recent Keck telescope observations
with the HIRES spectrograph (Vogt \etal 1994) show that the $b$
parameters are smaller and that the column density and $b$ parameters
distributions vary with redshift. In the ensuing sections, we consider
the effect on the observed colors by varying these input
distributions. The six specific cases are as follows:

\begin{itemize}

\item ``MLOS-EW:'' Madau's mean attenuation curve method with Lyman
forest clouds selected from an equivalent width distribution from
Murdoch \etal (1986), and Lyman limit clouds selected from a column
density distribution from Tytler (1987) and Sargent \etal (1989). For
each Lyman forest cloud 17 Lyman series lines were computed with
$b$ parameters constant at 35 km/s. The discrete absorber
distribution, as a function of equivalent width, $W$ and redshift,
$z$, is given by:

\begin{equation}
\frac{\partial^2 N}{\partial W \partial z} =
    \left\{ \begin{array}{l}
    \frac{3.4}{W_*} \ \left({\frac{W}{W_*}} \right)^{-\beta}
      \left( 1+z \right)^{\gamma}
      \qquad \left( W < 0.2~\hbox{\AA} \right) \\
    \frac{12.2}{W_*} \ e^{-W/W_*}\left( 1+z \right)^{\gamma}
      \qquad \left( W > 0.2~\hbox{\AA} \right) 
    \end{array} \right.,
\end{equation}

\noindent where $W_* = 0.3$ \AA, $\beta = 1.5$, and $\gamma = 2.46$.
This distribution was used to add the Lyman series lines to the
spectrum. However, in this ``MLOS-EW'' procedure, following Madau
(1995) and Madau \etal (1996), the Lyman limit breaks corresponding to
these clouds were added in by choosing $N_H$ from the distribution:

\begin{equation}
\frac{\partial^2 N}{\partial N_{HI} \partial z} =
    \left\{ \begin{array}{l}
    2.4~\hbox{x}~10^7 \ N_{HI}^{-1.5} \left( 1+z \right)^{2.46} \qquad
      \left( 2~\hbox{x}~10^{12} < N_{HI} < 1.59~\hbox{x}~10^{17}~\hbox{cm}^2
      \right) \\
    1.9~\hbox{x}~10^8 \ N_{HI}^{-1.5} \left( 1+z \right)^{0.68} \qquad
      \left( 1.59~\hbox{x}~10^{17} < N_{HI} < 1~\hbox{x}~10^{20}~\hbox{cm}^2
      \right)
    \end{array} \right.
\end{equation}

\noindent These equivalent width and column density distributions are not
entirely self--consistent, but are approximately so based on
a curve of growth analysis.  They were used for calculational 
convenience in this method.

\item ``MLOS-NH:'' Madau's mean attenuation curve method with Lyman
forest and Lyman limit clouds selected from column density
distributions above (equation 8). For each Lyman forest cloud 17 Lyman
series lines were computed with a $b$ parameter constant at 35
km/s. Unlike ``MLOS-EW'', this model uses the same self--consistent
distribution function to add the Lyman series and Lyman limit effects.
The curve of growth was used to derive the equivalent width from the
column densities chosen from this distribution.

\item ``MC-NH:'' Monte Carlo method with Lyman forest and Lyman limit
clouds selected from the same distributions as MLOS-NH. For each Lyman
forest cloud 17 Lyman series lines were computed with a $b$ parameter
constant at 35 km/s.  This is therefore the same model as MLOS-NH
(same distributions) except that it is applied via the Monte Carlo
method instead of the mean line-of-sight method.

\item ``MC-NHb:'' Monte Carlo method with Lyman forest and Lyman limit
clouds selected from equation 8. For each Lyman forest cloud 25 Lyman
series lines were computed with $b$ parameters selected from a
distribution independent of $N_H$ and $z$. The $b$ parameter
distribution is given by a Gaussian probability distribution centered
around 23 km s$^{-1}$, of width 8 km s$^{-1}$, which is truncated
below 15 km s$^{-1}$, as observed in the range $3.4 < z < 4.0$ by Lu
\etal (1996).  These lines are considerably narrower than those used
in the previous three distributions, and a comparison of this model to
``MC-NH'' should allow us to determine the magnitude of the effect of
changing the $b$ distribution on the colors.

\item ``MC-Kim:'' Monte Carlo method with Lyman forest and Lyman limit
clouds selected from redshift dependent column density distributions
estimated from Kim \etal (1997).  For each Lyman forest cloud 25 Lyman
series lines were computed with $b$ parameters selected from a
distribution (also from Kim \etal) dependent on $z$ but not on
$N_{HI}$. The discrete absorber distribution, as a function of column
density and redshift is given as:

\begin{equation}
\frac{\partial^2 N}{\partial N_{HI} \partial z} =
    \left\{ \begin{array}{l}
    3.14~\hbox{x}~10^7 N_{HI}^{-1.46} \left( 1+z \right)^{1.29} \qquad
      \left( 1~\hbox{x}~10^{12} < N_{HI} < 1~\hbox{x}~10^{14}~\hbox{cm}^2
      \right) \\
    1.7~\hbox{x}~10^6 N_{HI}^{-1.46} \left( 1+z \right)^{3.1} \qquad
      \left( 1~\hbox{x}~10^{14} < N_{HI} < 1.59~\hbox{x}~10^{17}~\hbox{cm}^2
      \right) \\
    1.9~\hbox{x}~10^8 N_{HI}^{-1.5} \left( 1+z \right)^{0.68} \qquad
      \left( 1.59~\hbox{x}~10^{17} < N_{HI} < 1~\hbox{x}~10^{20}~\hbox{cm}^2
      \right)
    \end{array} \right.,
\end{equation}

\noindent which is a rough approximation to the data points given in
Fig. 1 of Kim \etal (1997).  The Lyman limit system regime is still
best--described by the same observed distribution as above since the
HIRES/Keck sample is not large enough to consider this
population. Similarly the $b$ parameter distribution is given, again
very approximately from Kim \etal (1997) by a Gaussian probability
distribution centered around $-3.85~z_{gal} + 38.9$ km s$^{-1}$, of
width $-3.85~z_{gal} + 20.9$ km s$^{-1}$, which is truncated below
$-6.73~z_{gal} + 39.5$ km s$^{-1}$. This reflects a general trend for
the lines to become broader with decreasing $z$.  Kim \etal (1997) see
a slight tendency for this evolution to be more pronounced in higher
column density systems.  Since our goal is to give the range of colors
possible due to uncertainties in the observed distribution functions
of the Ly$\alpha$ forest clouds, and for simplicity, we opt to adopt
this $b$ parameter distribution which is independent of $N_{HI}$.
These model distributions can be refined when larger high resolution
data-sets become available.

\item ``MLOS-Kim:'' Madau's mean attenuation curve method
with Lyman forest and Lyman limit clouds selected from the same column
density and $b$ parameter distributions as MC-Kim above.

\end{itemize}

To summarize our models: MLOS-EW and MLOS-NH contrast two different
versions of distributions discussed by Madau (1995). MLOS-NH and MC-NH
are identical in terms of input distributions, and vary only by
method. MLOS-Kim and MC-Kim also vary only by method, but for a
different set of input distributions than MLOS-NH and MC-NH.  A
comparison of MC-NH and MC-NHb highlight the effects of adopting a
realistic $b$ parameter distribution. Finally, MC-NHb and MC-Kim
contrast the best possible input distributions prior to and after the
most recent HIRES/Keck observations of Kim \etal (1997).

\subsubsection{Spectra}

We adopt two different spectra in the ensuing analysis: (a) a flat
$f_\lambda$ spectrum and (b) a model galaxy spectral energy
distribution (SED) generated assuming a constant star formation rate
(CSFR), 0.1 Gyr age, Salpeter initial mass function in the mass range
$0.1 < M/M_{\odot} < 125$, and solar metallicity (Bruzual \& Charlot,
1993). To the second spectrum we have added internal reddening of
$E(B-V)=0.2$ using the effective dust attenuation curve for starburst
galaxies from Calzetti (1997). The choice of SED and reddening are
roughly representative of the expected and observed properties of
galaxies found at high redshifts (e.g.  Madau \etal 1996, Steidel
\etal 1996, Lowenthal \etal 1997, Pettini \etal 1997).\footnote{Note
that there are differences of order 0.1-0.3 mag in unattenuated color
between the Bruzual \& Charlot (1993) models used here and more recent
versions of the same, due to improvements in the stellar tracks and
libraries. However, these differences are small compared to the
effects of intervening absorption, and do not qualitatively change the
position of the redshift tracks with respect to the selection
boundaries of e.g.  Madau \etal (1996). For these reasons, the
specific choice of models is irrelevant to the analysis at hand.}  The
SED described above just so happens to be fully contained within the
$2 < z < 3.5$ selection box defined by Madau \etal (1996). We refer to
this SED as the ``CSFR'' spectrum.

The flat spectrum is used primarily to illustrate the distribution of
attenuation as a function of wavelength, and also to contrast with the
more realisted SED to demonstrate the importance of color terms.  We
use only a single realistic SED in our color distribution analysis.
While the results of our analysis will indeed be SED-dependent
(i.e. there exist 'color terms'), recall that the thrust here is to
pick a single SED representative of the high-redshift galaxy
population, and study in detail the effects of varying the neutral
hydrogen absorption.


\section{Comparison To Madau's Analytic Approximation}

%
%

\subsection{Mean Line of Sight}

As a final test of our Monte Carlo method, we show in Figure 3 that we
are able to reproduce the analytic estimates from Madau (1995, Figure
3) of the mean transmission function at $z = 3$ using the identical
absorber parameters. The agreement is remarkably good. The opacity
(i.e. 1 - transmission) between 1216-912 \AA \ in the galaxy
rest-frame (here $3648 < \lambda (\AA) < 4864 $) comes from
Lyman-series absorption of intervening neutral hydrogen clouds. Each
'step' in the transmission function represents the addition of a new
line in the series. The transmission function slopes upwards between
steps towards decreasing wavelength because this corresponds to
shifting to lower redshifts where absorbers are rarer at a given
column density.  For any individual line of sight, we note, these
steps are more difficult to discern because of the discrete nature of
the absorption.  Below 3648 \AA \ the opacity is dominated by the
Lyman limit from the same systems, in particular from the systems at
highest column density.  For the average line of sight the
transmission never recovers at shorter wavelengths (within the range
of these simulations). However, for an individual line of sight, the
transmission can sometimes recover, again because of the discrete
nature of the absorption. Such effects have been observed in real
lines of sight (e.g. Reimers \etal 1992).

Figure 3 also illustrates the effect of changing the column density
and $b$ parameter distributions on the mean LOS transmission. Below
the Lyman-break in the galaxy rest-frame (3650 \AA \ in this example
at $z \sim 3$), the differences in the resulting mean monochromatic
transmission are small, typically 10\% or less. However, at shorter
wavelengths, such as those sampled by $U_{300}$, the variations are
considerably larger. The largest variations are between the MLOS-EW
and MC-Kim models, which amount to $\sim 0.8$ mag at 3000 \AA \ at
this redshift. There is very little difference between the MC-Madau-NH
and MC-Madau-NHb models. This is as expected since the only difference
between these two models are are the $b$ parameters; below the Lyman
break the $b$ parameters are unimportant. The opposite is true in the
other regime: differences in the MLOS transmission in the Lyman
series region of the spectrum ($3700 \AA \ < \lambda < 4864 \AA$) are
most pronounced between MC-NH (or MLOS-NH) and MC-NHb which differ
only in their $b$ parameter distributions. This points to the
importance of including a realistic $b$ parameter distribution in the
models.  Here the MC-NHb and MC-Kim model transmission curves are
quite similar even though their column density distributions are
substantially different. The MC-NHb and MC-Kim models also have
different $b$ parameter distributions. From this we may surmise that
the precise formulation of the $b$ parameter distribution is not a
critical ingredient of the models {\it as long as some attempt is made
to provide a reasonably accurate representation,} such as we have done
in both the MC-NHb and MC-Kim models.  Specifically, a constant $b$
parameter of 35 km s$^{-1}$ will yield incorrect transmissions at the
10\% level.

More relevant for the color selection of high redshift galaxies is to
compare the differences in the attenuated magnitudes within broad
bands.  This is illustrated in Figure 4 for $U_{300}$ and $B_{450}$
bands over a range in redshift from $1.75<z<5$.  In the top panels the
MLOS-NH and MC-NH models are considered for the two input
spectra. Recall that both models use the same $N_{HI}$ and
$b$ parameter distributions. The largest differences in this set
of curves is between those using different input spectra.  This
illustrates the considerable effect of ``colors terms'' on the
magnitude increments.  The color terms are present in both bands, but
are particularly large ($> 2$ mag for $z>3.2$) in the $U_{300}$ band
due to the red leak of the F300W filter. In the redder bands,
$V_{606}$ and $I_{814}$, color terms are negligible over the range of
redshifts considered. However, the color terms in the $B_{450}$ band
are also quite considerable, reaching $\sim 1$ mag at $z\sim4.4$. {\it
Even though the transmission functions are independent of the
intrinsic spectrum, the attenuation in broad-band magnitudes are not.}
Hence mean transmission function cannot simply be convolved with
filter response curves to generate universal magnitude increments
for all galaxy spectra. For example, the magnitude increments in
Figure 4 of Madau (1995) or Figure 1 of Madau \etal (1996) are only
valid for the specific spectrum used in the calculation (in those
examples the effective input spectrum was flat in $f_\lambda$).

There are smaller differences present in the top panel of Figure 4
between the MLOS-NH and MC-NH models -- for a given input spectrum.
These differences represent the effect of taking logarithms at
incorrect or correct stages of calculating magnitude increments,
i.e. using the mean cosmic transmission (MLOS) or Monte Carlo
simulation (MC) methods, respectively.  In the MLOS method, the
magnitude increment is the log of the mean transmission, whereas the
mathematically correct calculation is to take the mean of the log of
the transmission of the individual lines of sight. In general, the two
are not mathematically equivalent. The effect is appreciable in
certain redshift ranges, particularly for the $U_{300}$ band near $z=3.5$,
where differences approach 1 mag. The effect is compounded when
calculating colors.

The bottom panels illustrate the combined effects of different methods
and extreme $N_{HI}$ and $b$ parameter distributions. Solid and
dashed curves are similar to the top panels, except here representing
the MC-Kim and MLOS-Kim models. The MLOS-Kim model is only included
here for the flat input spectrum.  Dotted curves are estimates of
magnitude increments using the mean transmission curve from the
Madau-EW distributions. Note the similarity between the MLOS-EW and
MC-Kim increments for the flat input spectrum, relative to MC-Kim and
MLOS-Kim increments.  Fortuitously, the differences in $N_{HI}$ and
$b$ parameter distributions between the MLOS-EW and MC-Kim models
cancel the effect of using the wrong calculation method for the
$U_{300}$ band for a flat $f_{\lambda}$ spectrum. However, significant
differences remain in the $U_{300}$ band for the CSFR spectrum and for
both spectra in $B_{450}$ band.

\subsection{Line of Sight Variations}

As discussed in \S2.1, the MLOS method is problematic not only for
estimating the mean colors of high redshift object, but also for
estimating the dispersion in these colors (for a given redshift and
input spectrum).  In Figure 5, we reproduce the mean transmission
function at $z=3$ for the MLOS-NH model via our Monte Carlo method
(MC-NH). This is done simply by averaging, wavelength by wavelength,
the transmission functions for 4000 individual lines of sight. We can
then also directly calculate the standard deviation about the mean
transmission function (i.e. mean $\pm\sigma$), also illustrated in
Figure 5.  Calculating standard deviations in this way results in
unphysical values for the transmission function, i.e. above 1 and
below 0.  Even truncating at these limits, these curves substantially
over-estimate the rms scatter, as discussed in \S2.

The Monte Carlo method permits us to calculate reliably the true
dispersion in line-of-sight attenuation as a function of band-pass and
redshift.  In Figure 6 the dispersion in magnitude increments is
plotted for the $U_{300}$ and $B_{450}$ bands as a function of
redshift for the CSFR spectrum and the MLOS-NH and MC-NH models. (As
before, the results {\it do} depend on the input spectrum due to color
terms; the CSFR spectrum is most representative for real galaxies.)
The mean magnitude increments are for the MC-NH model (top panels,
Figure 4), i.e. they correctly are the mean of the magnitude
increments over an ensemble of lines-of-sight. The dispersion is shown
for the MC-NH model (about this mean), and for the MLOS-NH model. The
MLOS-NH dispersion is based on the standard deviation about the mean
of the transmission curve. The MC-NH dispersion curves represent the
$67^{th}$ percentiles above and below the mean, {\it as calculated
from the distribution of magnitude increments.}  In the $U_{300}$
band, the dispersion estimated via the MC-NH model is typically much
less than the $\pm 1 \sigma$ dispersion in magnitude increments
calculated from MLOS-NH, becoming comparable only at $z>3.5$ where the
filter red-leak begins to dominate the detected flux.  For the
$B_{450}$ band the MC-NH dispersion is typically much less than a
tenth of the MLOS-NH dispersion. Over the redshift range of interest,
i.e. $2<z<3.5$, the $B_{450}$ dispersion is more highly overestimated
by MLOS-NH compared to MC-NH than is the $U_{300}$ band dispersion.

\subsection{The apparent colors of a constant star-forming galaxy at $z=3$}

We can now fold together the above results to show the magnitude of
the systematic errors inherent in the MLOS-type calculations for the
estimation of the distribution of high redshift galaxy colors. In
Figures 7a and 7b, the two-color diagrams used for selecting
high-redshift galaxies in the Hubble Deep Field (Madau \etal 1996) are
depicted for a single galaxy spectral type at $z=3$ and $z=4.5$,
respectively. The galaxy spectral type corresponds to the CSFR
spectrum. The redshifts were chosen to place Ly-$\alpha$ and the Lyman
break at comprable wavelengths with respect to the red sides of the
$U_{300}$ and $B_{450}$ filters at $z=3$ and the $B_{450}$ and
$V_{606}$ filters at $z=4.5$. The best estimate for the mean and
distribution of colors is represented by the open-star and small dots,
respectively, derived using the MC-Kim model for 1000 lines of
sight. Were one to use the MLOS-NH model (filled circles), the mean
color would be off by $\sim0.1$ mag. As we will see in the next
section, this systematic error increases substantially with redshift.
These differences are a confluence of different $N_{HI}$ and $b$
parameter distributions, and the mathematically incorrect procedure of
taking logs of means.

Even more spectacular is the difference in the predicted distribution
of colors between the MC-Kim and MLOS-NH models in three fundamental
characteristics.  Compared to the MC-Kim predictions, the MLOS-NH
method (i) substantially over-predicts the range of colors,
particularly the redder color ($(B_{450}-I_{814})$ in Figure 7a and
$(V_{606}-I_{814})$ in Figure 7b); (ii) yields marginal distributions
of colors which are skewed, i.e.  the implied 1$\sigma$ range is
larger towards the red than the blue about the mean for all colors;
(iii) predicts color distributions which are substantially misaligned
in color space. This last difference is especially important because
unlike the MC-Kim model, the MLOS-NH model yields apparent color
distributions at a {\it single} redshift that tend to lie along the
redshift trajectory in color space. Similar differences occur between
all MC-type and MLOS-type models. These differences have significant
ramifications for the selection efficiency of high redshift galaxies,
as discussed in \S5.

As with the mean, the differences in scatter between MC-Kim and
MLOS-NH models are a confluence of different $N_{HI}$ and $b$
parameter distributions and the mathematically incorrect procedure of
taking logs of means (MLOS methods). With the scatter, however, there
are two additional effects. First, the amplitude of the scatter and
the detailed distribution of colors is overestimated with the MLOS
method because dispersion in the line-of-sight attenuation was
incorrectly estimated at high spectral resolution (\S2.2). Second, the
mis-alignment of the scatter in colors is due to the fact that in the
MLOS methods there is a high degree of covariance between the
magnitude increments in different bands. This is due to the
transmission function shape being self-similar from one line of sight
to the next. In the MC method, the transmission function has a
different shape for each line of sight because the intervening
absorption is modeled as discrete, random process.

The only other effect to consider is color terms. This substantial
effect is illustrated in Figures 7a and 7b by considering what would
happen were we to take the unattenuated colors of a galaxy with the
CSFR spectrum (marked 'x') and apply magnitude increments based on the
MLOS-NH model using a {\it flat} input spectrum. The resulting
'attenuated' colors are illustrated by the arrows in these
figures. The error in the mean color now becomes quite substantial: $>
1$ mag in the bluer color ($(U_{300}-B_{450})$ or $(B_{450}-V_{606})$),
and about twice that of the previous error associated with MLOS-NH
using the right spectrum for the redder color ($(B_{450}-I_{814})$ or
$(V_{606}-I_{814})$).  Such large differences clearly will affect the
selection process of high redshift galaxies, particularly near the
selection boundaries in color space.


\section{The Distribution of Galaxy Colors From $1.75<z<4.5$}

We now explore the effects on the color distribution of galaxies as a
function of redshift when the attenuation models are varied. We
consider variations due both to method (MLOS vs MC) and to different
parameter distributions. These effects are summarized in Figures 8a
and 8b. For discussion purposes, we focus on differences that pertain
to redshifts within or near the boundary of the selection boxes
defined by Madau \etal (1996) to locate high redshift galaxies. These
redshifts are germane to the discussion of selection functions in the
next section. Also note that the wide range of colors at a given
redshift are due to line-of-sight variations and intervening-absorber
model differences; recall we are considering here only a single input
spectrum, i.e the CSFR galaxy spectrum.

Our best estimates for the mean and range of colors (illustrated in
these figures by the grey points) correspond to the MC-Kim model.  At
higher redshifts ($z>3.25$), the distribution in Figure 8a develops a
truncated upper limit in $(U_{300}-B_{450})$. This corresponds to the
redshift where the scatter in the magnitude increments for the CSFR
galaxy spectrum starts precipitously decreasing (Figure 6). What is
happening is that by this redshift the attenuation at $\lambda < 4000$
\AA \ (observed) has become so large that the flux in $U_{300}$ band is
dominated by the red leak where there is little intervening
absorption. Hence the scatter in observed $(U_{300}-B_{450})$ color is
dominated by the scatter in the $B_{450}$ magnitude increment.
Similarly, because there is little attenuation at these redshifts in
the $I_{814}$ band, the scatter in the $B_{450}$ magnitude increment
also dominates the scatter in the observed $(B_{450}-I_{814})$
color. Hence at the highest redshifts in Figure 8a, the color
distribution tends towards a line of slope -1 in the
$(U_{300}-B_{450})$ vs $(B_{450}-I_{814})$ two-color diagram, as
expected.  This behavior is not observed in Figure 8b for two
reasons. First, at the highest redshift plotted (4.75), the Lyman
limit has not moved completely through the $B_{450}$ bandpass. Second,
the $B_{450}$ does not have an appreciable red leak. By $z=5$, galaxies
along the most attenuated lines of sight will start completely
dropping out of the $B_{450}$ band and the $(B_{450}-V_{606})$ color
will become undefined.

The MC-NHb model, which represents the best $N_{HI}$ and $b$ parameter
distributions available prior to recent Keck observations, yields mean
colors which differ only slightly from the MC-Kim model means for
$z\leq3$. Indeed, we have checked that the two-color distributions in
these figures have similar scatter (shape and amplitude) over the
range $1.75<z<3.75$ for $(U_{300}-B_{450})$ vs $(B_{450}-I_{814})$ and
over the range $2.75<z<4.75$ for $(B_{450}-V_{606})$ vs $(V_{606}-I_{814})$.
Consequently we display MC-Kim simulation points and just the mean of
the MC-NHb distributions.

Figures 7 and 8 reveal MC-NHb compares more favorably with MC-Kim at
$z=3$ than does MLOS-NH. This reflects both the improved method (Monte
Carlo), plus the addition of a realistic $b$ parameter
distribution. (Recall that MLOS-NH assumes a constant $b$ parameter of
35 km s$^{-1}$. Additional discussion of the relative importance of the
$b$ parameter in estimating attenuation due to intervening clouds can
be found in Giallongo \& Trevese 1990.)  However, at higher redshifts
the MC-NHb model mean colors become systematically redder with respect
to MC-Kim model colors. This is most noticeable in the Figures 8a and
8b for the redder color because of the scale. At $z=4.5$, for example,
the mean $(V_{606}-I_{814})$ color for the MC-NHb model are shifted
redwards to the mean value for the MC-Kim at $4.625$. A similar
comparison of figures 7 and 8 also reveals that the MLOS-EW fares
worse (with respect to MC-Kim) than MLOS-NH. By $z=4$, the MLOS-EW
colors are comparable to the MC-Kim colors (in the mean) at $z =
4.125$, and by $z=4.5$, the jump has increased to 0.25 in $z$.

Note that while it is our expectation that the MC-Kim model is likely
to be the best estimate of the true distribution of line-of-sight
attenuation, it is not necessarily correct.  The two models MC-NHb and
MC-Kim can be considered illustrative of the possible systematic
errors in estimating high redshift galaxy colors due to our imperfect
knowledge of intervening absorption. Likewise, difference between the
MC-Kim and MLOS-EW models can be considered the worst-case scenario
were one to use the wrong method and absorber properties.

In summary, the predicted mean colors are very similar between models
for $z<2.5$, where intervening absorption is small ($<0.2$ mag for the
most sensitive color, $(U_{300}-B_{450})$). In an intermediate range
$2.5<z<3.5$, the mean colors compare reasonably well for MC-NHb and
MC-Kim, but in the worse-case comparison, MC-Kim vs MLOW-EW, mean
color differences are becoming substantial ($>0.5$ mag for the most
sensitive color, $(U_{300}-B_{450})$). In the highest range, $z>3.5$,
the mean color differences continue to grow, exceeding 1 mag in
$(B_{450}-V_{606})$ by $z=4.5$, and exceeding 0.5 mag in the reddest
color, $(V_{606}-I_{814})$, by $z=4.75$.  The distributions in color
for all redshifts are highly skewed for the MLOS-EW method relative to
both MC-Kim and MC-NHb (which are themselves comparable).  The sense
of the skewed color distribution places the scatter roughly along the
redshift trajectory in color space, particularly in
$(U_{300}-B_{450})$ vs $(B_{450}-I_{814})$. This results has direct
bearing on the selection of high redshift galaxies.


\section{Application to Selection Completeness}

To date, the selection of high redshift galaxy candidates in two-color
diagrams has been based on fixed, discrete boundaries in color.
Indeed, a similar situation prevails for the multi-color selection of
QSOs. For this reason, line-of-sight variations frequently will lead
to incompleteness in the true selected fraction of sources within some
finite range of redshift. Incompleteness increases for redshifts and
spectra where the mean color (over all lines-of-sight) is close to a
selection boundary. However, this will occur at different redshifts
for different intrinsic (unattenuated) SEDs.  For this reason, we
simplify the analysis to the consideration of the single CSFR spectrum
as a function of redshift.

The incompleteness will also be larger, at a given redshift, when the
mean color approaches a selection boundary perpendicular to the
direction in which the line-of-sight variations in color are
greatest. Recall from the results of the previous sections that the
variations tend to be substantially larger in the bluer color of a
two-color diagram. Hence the boundaries in blue colors, i.e. where the
redder color is not held constant, is where the most substantial
incompleteness (and consequently, contamination) might be
expected. (This corresponds to the lower boundaries in
$(U_{300}-B_{450})$ in Figure 8a and $(B_{450}-V_{606})$ in Figures
8b.)  Ameliorating this effect is the fact that colors tend to be near
such boundaries only at relatively low redshifts where the variations
are smaller -- at least for the specific choice of SED shown in
Figures 8a and 8b. However, given the very wide possible range of
intrinsic SEDs, e.g. Madau \etal (1996, see their Figures 3 and 5),
there may be sources which approach the blue-color boundaries at
higher redshifts where the variations are larger.

The specific selection function for the CSFR spectrum and the
selection boundaries in Figures 8a and 8b are shown in Figures 9a and
9b, respectively. These figures illustrate the effects of
line-of-sight variations in color as well as differences between
models for the intervening absorption.  Were there no line-of-sight
variations, the selected function would be a perfect square-wave
function of redshift, with the precise cut-on and cut-off redshifts
depending on the specific models.  Since our best model, MC-Kim,
yields mean colors that tend to be bluer than the other models, the
cut-on and cut-off redshifts are generally the highest. Intercomparing
the Monte-Carlo models only, the difference in redshift can be as
large as $\Delta z \sim 0.2$. The differences are more extreme when
comparing MC-Kim to MLOS-EW, but this is due not only to differences
in the mean colors but also to differences in the variance (scatter)
in the colors at a given redshift.

Line-of-sight variations in colors effect the selection function
equivalently to passing a smoothing filter over the square-wave
function that would be determined in the absence of scatter.  The
effect for the MC-type models is seen within about $\Delta z \sim
0.15$ of cut-on and cut-off redshifts, and is most extreme at the
cut-on redshift for the $(B_{450}-V_{606})$ vs $(V_{606}-I_{814})$
selection. However, this pales compared to the effect of the scatter
in the MLOS-EW model, which affects the selection function over
$\Delta z \sim 0.5$, which is equivalent to the half width of the
selection function in redshift. Conveniently, this model for the
intervening absorption is incorrect.

While the effects of line-of-sight variations in color do not
radically transform the selection for the MC-type models from a
square-wave, it is important to note that this result is for {\it a
single SED.} In comparison, the more complex selection functions in
Madau \etal's (1996) Figures 4 and 6 are due to variations in SEDs; in
their analysis {\it no intrinsic scatter} was assumed. To assess the
relative importance of line-of-sight variations vs intrinsic
differences in colors requires some {\it a priori} knowledge of the
range and distribution of SED types at high redshift.  At this time
this information is not available. The selection functions in Figure
9a and 9b here represent our current best estimate for the specific
CSFR spectrum. The method can be applied to any spectrum {\it on an
individual basis.}

A related point of note is that for any of our MC-type models, there
is a gap in redshift of $\Delta z \sim 0.25$ where neither two-color
selection boundary detects our example SED (CSFR spectrum) above the
20\% level. For the MC-Kim model this range is $3.4<z<3.65$.  Similar
gaps undoubtedly exist for different SED types, shifted appropriately
in redshift due to differences in intrinsic color. In the selection
function plots of Madau \etal (1996), no such gap is apparent. This is
illusory; again, this is because these selection functions represent
an average over many SED types.

Indeed, for all of these reasons it is not possible to generalize the
specific results in this section to a global analysis of the selection
function for high redshift objects. The only correct approach is to
consider the selection function for individual SEDs.  The implication
is that space densities must be calculated as a function of spectral
type, but this is astrophysically of interest anyway.


\section{The Effects of Internal Absorption}

So far we have only considered the effects of attenuation due to
intervening neutral hydrogen, and we have ignored the possibly
considerable effects of neutral hydrogen within (or immediately
surrounding) the high redshift galaxy or QSO. Depending on the column
density of this internal neutral hydrogen, it may well dominate the
attenuation since the Lyman break in the source rest-frame will
attenuate the {\it largest} range in observed (redshifted) wavelength.
Variations in the line-widths of optically-thick local gas could also
produce variations in the observed attenuation.

Considering only high redshift galaxies for the moment, what is the
expected column density of internal neutral hydrogen as observed for a
particular line of sight? As observed for the Milky Way from the
vantage point of our solar system, the internal neutral hydrogen
column density is not uniform. The lowest column density observed is
the ``Lockman Window'' (Lockman \etal 1986), where $N_{HI} \sim
4.5\times10^{19}$ cm$^2$. However, the solar system is in a
particularly neutral hydrogen-rich location within the galaxy. Stars
higher above the plane of the disk or at larger or smaller
Galacto-centric radii (even at the same height above the disk)
undoubtedly have sight lines at substantially lower column densities.

What we would really like to know is what is the effective attenuation
of high redshift galaxies to out-going UV ionizing photons.
Unfortunately, the effective attenuation of even the Milky Way is
unknown, and hence it is difficult to reliably speculate about the
effective attenuation for any other galaxies, let alone those at high
redshift. Direct upper limits exist for only four nearby galaxies
(Leitherer \etal 1995), based on HUT observations.  Deharveng \etal
(1997) estimate that the mean effective attenuation to Lyman continuum
photons in nearby galaxies should be $< 1$\%, but this value is
uncertain. Nonetheless, we do know there are substantial variations in
$N_{HI}$ along different sight lines out of the Milky Way, and we also
know from quasar pairs that absorption line systems have non-uniform
$N_{HI}$. It is therefore not unreasonable to expect that whatever
effective attenuation is present in high redshift galaxies, it may
vary widely for a single galaxy depending on the line of sight (from
the galaxy's perspective). This is consistent with results from
studies of how ionizing flux escapes from star forming regions in
galaxies (e.g. Dove \& Shull 1994, Patel \& Wilson 1995a,b, and
Ferguson \etal 1996). Consequently it is likely that the effective
attenuation varies substantially from one galaxy to another. Since the
lines of sight at the lowest column density will dominate the UV flux,
it is conceivable that the effective attenuation may {\it appear} to
be quite low.

In Figures 10a and 10b we repeat the experiment of calculating the
color distribution of the CSFR spectrum over a wide range in redshift
between $1.75<z<5$ using the MC-Kim model for intervening absorption.
To this model we have added (spatially uniform) internal column
densities of $10^{17}, 10^{18}$, and $10^{19}$ cm$^2$. The MC-Kim
model with no internal $N_{HI}$ absorption is included for reference.
The first thing to note is that for an internal column density of
$10^{17}$ cm$^2$ the range and mean of the colors are only moderately
different from the case where there is no internal absorption.  Below
$10^{17}$ cm$^2$, there is little effect on the mean or distribution
of colors. The largest change in the color distribution is between
$10^{17}$ and $10^{18}$ cm$^2$. However, above $10^{18}$ cm$^2$ there
is virtually no change in the observed distribution of colors. In
summary, the critical value for the internal $N_{HI}$, in terms of
affecting the observed colors, is $\sim 10^{17.5}$ cm$^2$. Any column
density substantially in excess of $10^{17.5}$ cm$^2$ yield a color
distribution well approximated by the intervening absorption model
(here MC-Kim) plus $10^{18}$ cm$^2$ of internal absorption; any column
density substantially less than $10^{17.5}$ cm$^2$ yields a color
distribution well approximated by the intervening absorption model
alone.

The sense of the change in the mean and distribution of colors as a
function of increasing internal $N_{HI}$ is two-fold. First, the mean
color for the bluer color becomes redder, e.g.  $(U_{300}-B_{450})$ in
Figure 10a. Second, the range of colors for the bluer color becomes
smaller. The (decreased) range in color is roughly constant with
redshift, unlike the range in color where there is no internal
$N_{HI}$.  The near-constant range in the internally attenuated colors
is due to a canceling of two competing effects: with increasing
redshift, more of the bluest bandpass gets blocked out by Lyman break
from the internal neutral hydrogen, while the flux dispersion in the
remaining part of the bandpass increases (the color range with no
internal absorption increases). The two effects cancel to first order
until the Lyman break from the internal $N_{HI}$ sweeps through the
bluest bandpass ($z\sim3.4$ for $U_{300}$ and $z\sim4.9$ for
$B_{4500}$).

Note, however, that while the mean and distribution of bluer colors
changes with increasing internal $N_{HI}$, the redder colors in Figures
10a and 10b do not. This is because in the redder bands for the
redshifts displayed, the absorption produced by the internal neutral
hydrogen's discrete Lyman series lines does not compete with the forest
of Lyman series lines integrated along the sight-line, i.e. at
different redshifts or different observed wavelengths. 

The fact that the redder colors do not change has important
implications for the selection function. The selection functions
derived in the previous section (Figures 9a and 9b) are {\it virtually
unchanged} for our particular choice of SED (the CSFR spectrum). This
is because internal absorption does not appreciably affect the scatter
in the bluer color at low redshift where the selection boundary is
perpendicular to this color. At the high redshift limit of the
selection function, what is relevant is the scatter in the redder
color, but this is largely unchanged by the internal
absorption. However, had we chosen an intrinsically bluer SED such
that the mean colors were closer to the blue-color selection boundary
at higher redshifts, internal absorption {\it would} have substantial
effects on the selection function.  Because of the reduced range
(scatter) in the bluer color, the selection function would persist as
close to a square-wave function for a larger range of intrinsic galaxy
colors. While this may seem advantageous, it begs the question of what
the internal absorption really is. Indeed, for intrinsically very blue
SEDs whose mean colors approach the blue-color boundary at higher
redshifts, it is not possible the accurately determine the selection
function without detailed knowledge of the internal absorption.

\section{Summary}

We have shown that the attenuated colors of high redshift galaxies and
QSOs depends sensitively on method used to calculate this attenuation.
In order of importance, the mean and scatter in the colors is
sensitive to:

\begin{itemize}

\item The mathematical procedure, i.e. Monte Carlo (MC) vs mean
line-of-sight (MLOS). The Monte Carlo (MC) method is the only correct
way to estimate attenuated colors of high redshift objects; the mean
line of sight method (MLOS) results in substantial systematic errors
in both the mean and scatter of these colors.

\item The adopted distribution of intervening cloud column densities 
and $b$ parameters.  Our best estimate for the intervening cloud
column densities and $b$ parameters is represented by the MC-Kim
model, and yields significantly different colors than the best
previous estimates (e.g. those parameters used in the MC-NHb model).

\item The amount of internal absorption due to neutral hydrogen.  We
find that column densities less than $10^{17}$ cm$^2$ are unimportant,
while column densities above $10^{18}$ cm$^2$ all result in a similar
reddening of the bluer colors and reduction in their range at a given
redshift.

\end{itemize}

We have ignored, in this analysis, the variations in attenuation due
to dust. Instead, we have focused on the distribution of colors for a
single SED corresponding to constant star forming galaxy of 0.1 Gyr
age with fixed internal reddening of $E(B-V)=0.2$. This spectrum
should be representative of high redshift galaxies, but a much wider
range of spectral types are observed at high redshift (Pettini \etal
1997). Even within this simplified context the systematic differences
between models parameters and methods produce substantial differences
in the selection function as derived from fixed regions in two-color
diagrams presented by Madau \etal (1996). Variations in attenuation
due to dust will further complicate the situation, and need also to 
be modeled in a Monte Carlo fashion.

While internal neutral hydrogen absorption acts to diminish the
scatter in the bluer colors, it has little effect on the selection
function in most cases. If, however, the intrinsic color of an object
(i.e. its unattenuated color) results in a mean attenuated colors near
the blue boundary of a two-color selection regions, internal
absorption will result in a sharper selection function.  That is,
either the source will more uniformly selected or rejected as a high
redshift candidate.  For SEDs satisfying this criterion, the
difficulty in defining their selection function is due primarily to
our lack of knowledge of the effective attenuation from internal
neutral hydrogen in high redshift galaxies.

Because color-terms are important in the estimate of the attenuated
colors, it is not possible to generalize the results for any single
SED.  Even ignoring color terms, the selection function will be
different for each SED. As a result, a global selection function
determined for a distriution of model SEDs will only be correct in so
far as this distribution and the model SEDs reflect reality. This
distribution is {\it a priori} unknown. Hence the only correct way to
estimate the selection function for an ensemble of high redshift
objects, we conclude, is to do so for individual spectral types.

\acknowledgements We would like to thank Piero Madau for providing
encouraging comments on an initial presentation of this work at the
AAS, B. Wakker and J. Mathis for useful discussion, Chris Churchill
for assistance with the Ly$\alpha$ forest simulations, W. R. Weimer
for consultation with JMG on computational methods, and the anonymous
referee for helpful comments which deepened the discussion. MAB
acknowledges support from NASA grant NAG5-6032. JCC acknowledges
support from NASA grant NAG5-6399 and NSF AST--9617185.

\clearpage

\figcaption[]{The response curves for the $U_{300}$ (solid), $B_{450}$ 
(dashed), $V_{606}$ (solid), and $I_{814}$ (dashed) bands,
superimposed on a normalized galaxy spectrum at a redshift of 3.  The
unattenuated galaxy spectrum, described in the text, is represented by
the solid, bold line, while the attenuated galaxy spectrum for a
single line of sight (LOWRES Monte Carlo model) is in grey. Note the
red leak of the $U_{300}$ filter pealing at $\sim$7150 \AA.
\label{fig1}}

\figcaption[]{A portion of a simulated galaxy spectrum ($f_\lambda \ vs \
\lambda$), as observed at $z = 3$, for a single (representative) line 
of sight, attenuated by intervening absorption as described by the
MC-Kim model via the HIGHRES method (top panel) and the LOWRES method
(middle panel). The two models are compared in the bottom panel over a
30 \AA \ region. See text for description of methods. \label{fig2}}

\figcaption[]{The mean cosmic transmission functions (i.e. the average
line-of-sight, top panel) at $z=3$, in order from bottom to top:
MLOS-EW (short dashed line), MLOS-NH (heavy solid line), MC-NH
(light dotted line), MC-NHb (light solid line), MC-Kim (heavy
dotted line). The analytic transmission functions appear smooth
while the Monte Carlo simulations (over 4000 lines of sight) show
small statistical variations.  Note the MLOS-NH and MC-NH are
essentially identical on average. This is expected, and demonstrates
the accuracy of our Monte Carlo method. Also notice the difference
between the various models due to differences in adopted $N_{HI}$ and
Doppler $b$ parameter distributions. The most extreme cases are MLOS-EW
(Madau 1995, Madau \etal 1997) and MC-Kim (our best model).  The
average transmission functions at $z=3$ is displayed in the bottom
panel, scaled to units of monochromatic magnitudes. \label{fig3}}

\figcaption[]{Mean attenuation, in magnitudes ('magnitude increments') 
for F300W ($\Delta U_{300}$, left panels) and F450W ($\Delta B_{450}$,
right panels) passbands, derived in a variety of ways. In all panels,
magnitude increments are calculated for the flat and CSFR spectra
described in \S2.3.2.  In the top panels, dashed curves correspond to
estimating magnitude increments from the mean cosmic transmission
curve (MLOS-NH, Madau 1995, Madau \etal 1997), while the solid curves
represent the correct method of averaging the magnitude increments of
the Monte Carlo simulations (MC-NH). Both use the same $N_{HI}$ and
Doppler $b$ parameter distributions. Note the large differences between
methods (for the same spectrum) and spectra (for the same method),
illustrating the effects of procedural differences and color terms in
calculating magnitude increments. The bottom panels illustrate the
combined effects of different methods and extreme $N_{HI}$ and
Doppler $b$ parameter distributions. Solid and dashed curves are similar
to the top panels, except here representing the MC-Kim and MLOS-Kim
models, respectively. Dotted curves are estimates of magnitude
increments using the mean transmission curve from the MLOS-EW
distributions. Note significant differences in the $U_{300}$ band for
the CSFR spectrum and for both spectra in $B_{450}$
band. \label{fig4}}

\figcaption[]{The mean cosmic transmission function at $z=3$ (middle
curve) together with the $\pm 1 \sigma$ rms scatter caused by
statistical fluctuations in the number of absorbers along the path as
calculated using the formalism of Madau (1995) (top and bottom
curves). These curves are calculated from the usual moments (mean and
standard deviation) of an ensemble of transmission curves representing
many individual lines of sight using the MC-NH model. Note how the
$\pm 1 \sigma$ rms scatter exceeds the physical limits of 0 and 1 for
transmission (indicated by dashed lines). \label{fig5}}

\figcaption[]{Mean and dispersion of attenuation, in magnitudes 
('magnitude increments') for F300W ($\Delta U_{300}$, top panel) and
F450W ($\Delta B_{450}$, bottom panel) passbands. Mean attenuation for
the MC-NH model is shown as a solid line. Dispersions are calculated
in two ways: (1) $\pm$ 67th-percentile attenuation about the mean
attenuation for the MC-NH model (dotted curve), (2) attenuation
calculated from the $\pm 1 \sigma$ deviation about the mean
transmission function, i.e. the MLOS-NH method (dashed curve).
Method (1) is correct. \label{fig6}}

\figcaption[]{(a) The distribution of colors in $(U_{300}-B_{450})$
vs. $(B_{450}-I_{814})$ for a single galaxy spectrum (CSFR) as
observed at $z=3$ for (1) 1000 lines of sight (small dots) assuming
the MC-Kim model; (2) the mean color of these 1000 lines of sight (our
best estimate, star ); (3) the mean and $\pm$1 sigma attenuated colors
based on mean and $\pm 1 \sigma$ transmission functions using the
MLOS-NH model (filled circles connected with dashed lines); (4) the
unattenuated color (x); (5) the mean attenuated color based on the
unattenuated color and magnitude increments from Figure 4 calculated
from the mean transmission function and a flat spectrum for the
MLOS-NH model (open circle). Arrows, representing attenuation vectors,
connect point (4) with (2), (3) and (5). In addition to the large
offsets between various estimates of mean colors at $z=3$ for this
single SED, note the large difference in scatter in galaxy colors
between (1) and (3). \label{fig7a}}

\setcounter{figure}{6}

\figcaption[]{(b) The distribution colors in $(B_{450}-V_{606})$
vs. $(V_{606}-I_{814})$ for a single galaxy spectrum (CSFR) as observed
at $z=4.5$. Models and symbols are identical to Figure 7a, except the
star (mean color of the MC-Kim simulations) is shaded grey here for
visibility. \label{fig7b}}

\figcaption[]{(a) The distribution colors in $(U_{300}-B_{450})$
vs. $(B_{450}-I_{814})$ for a single galaxy spectrum (CSFR) observed
along random lines of sight at discrete redshifts between $1.75<z<4.5$
at intervals of 0.125 in $z$: (1) MC-Kim model for 4000 lines of sight
(grey points); (2) the mean colors for the MC-NHb model (open stars);
(3) the mean colors for the MLOS-EW model (open or filled squares),
with their $\pm 1 \sigma$ dispersions (thick, solid lines).  The
redshift track for the unattenuated colors is shown as a think, solid
line.  Redshifts are marked at $z=2$ and 2.5 (centered on the MC-Kim
distributions in $(U_{300}-B_{450})$), and $z=3$ and 3.5 (centered on
$(B_{450}-I_{814})$. At these redshifts the MLOS-EW points are filled
squares, and the unattenuated colors are open triangles. The selection
criterion of Madau \etal (1996) is represented by heavy, solid lines.
Galaxies with colors up and to the left of the solid lines would be
selected as high redshift galaxies by Madau \etal
(1996). \label{fig8a}}

\setcounter{figure}{7}

\figcaption[]{(b) The distribution colors in $(B_{450}-V_{606})$
vs. $(V_{606}-I_{814})$ for a single galaxy spectrum (CSFR) observed
along random lines of sight at discrete redshifts between $1.75<z<4.5$
at intervals of 0.125 in $z$.  Models and symbols are identical to
Figure 8a. Galaxies with colors inside the solid lines would be
selected as high redshift galaxies by Madau \etal (1996).
\label{fig8b}} 

\figcaption[]{(a) Selection functions for different attenuation models
but for a {\it single} SED: the CSFR spectrum. The selection functions
are defined by the selection boundary in the $(U_{300}-B_{450})$
vs. $(B_{450}-I_{814})$ two-color diagram of Figure 8a, adopted from
Madau \etal (1996), Figure 3. The specific models are
labeled. \label{fig9a}}

\setcounter{figure}{8} 

\figcaption[]{(b) Selection functions, as in Figure 9a, except defined
by the boundary in the $(B_{450}-V_{606})$ vs. $(V_{606}-I_{814})$
two-color diagram of Figure 8b, adopted from Madau \etal (1996),
Figure 5. \label{fig9b}}

\figcaption[]{(a) The distribution colors in $(U_{300}-B_{450})$
vs. $(B_{450}-I_{814})$ for a single galaxy spectrum (CSFR) observed
along random lines of sight at discrete redshifts between $1.75<z<4.5$
at intervals of 0.125 in $z$.  Redshifts of 2, 2.5, 3, and 3.5 are
labelled.  The MC-Kim model is considered only, but for three column
densities of internal neutral hydrogen: none (large, light-grey dots),
$10^{17}$ cm$^2$ (medium grey dots), and $10^{18}$ cm$^2$ (dark, small
dots). The distribution for $10^{19}$ cm$^2$ of internal neutral
hydrogen (not shown) is virtually identical to the $10^{18}$ cm$^2$
case. \label{fig10a}}

\setcounter{figure}{9} 

\figcaption[]{(b) The distribution colors in $(B_{450}-V_{606})$ 
vs. $(V_{606}-I_{814})$; models and symbols are as in Figure
10a. Redshifts of 3.5, 4, 4.5 are labelled. \label{fig10b}}

\setcounter{figure}{0} 

\clearpage
\begin{figure}
\plotfiddle{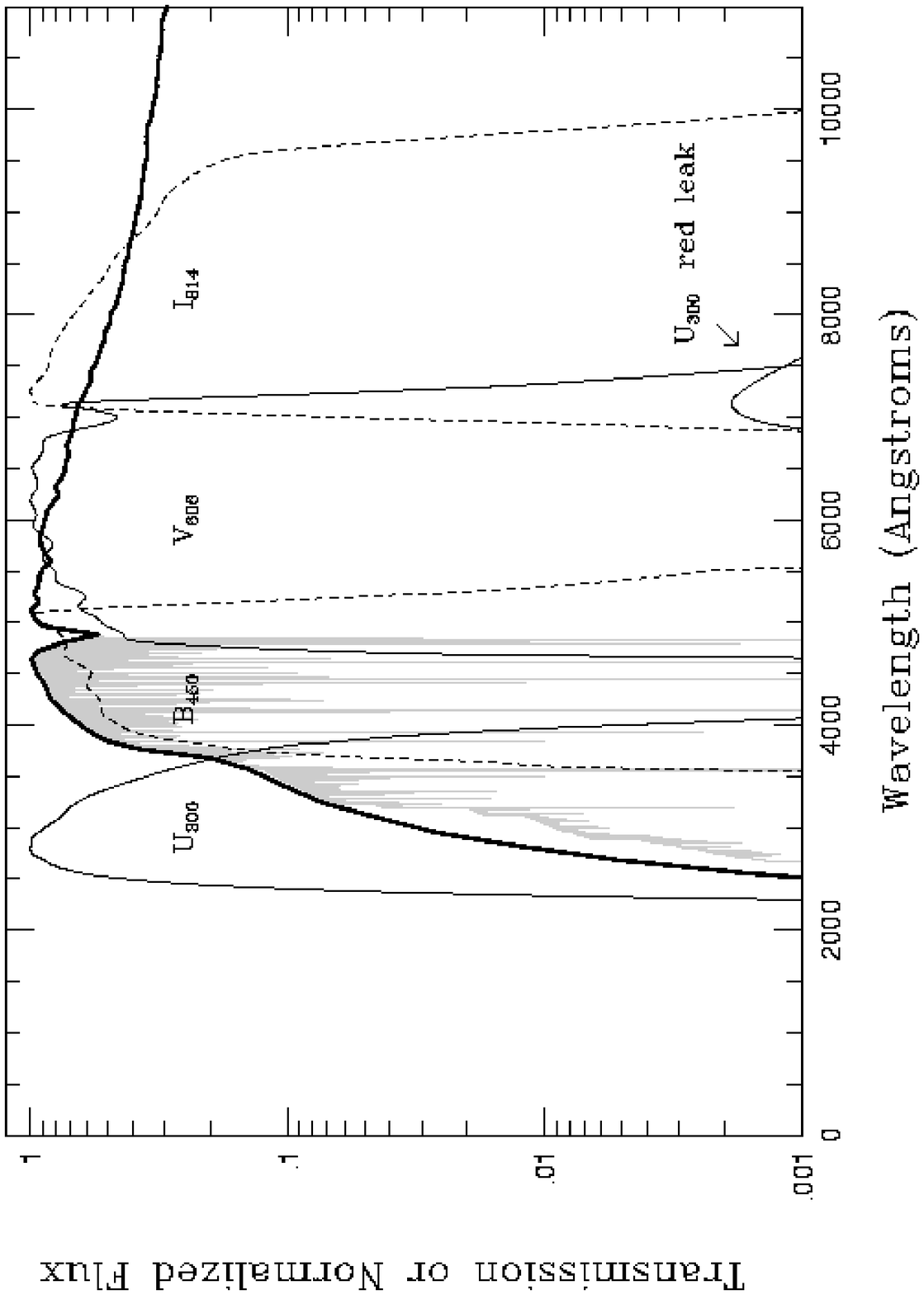}{8in}{0}{85}{85}{-250}{-25}
\caption{}
\end{figure}

\clearpage
\begin{figure}
\plotfiddle{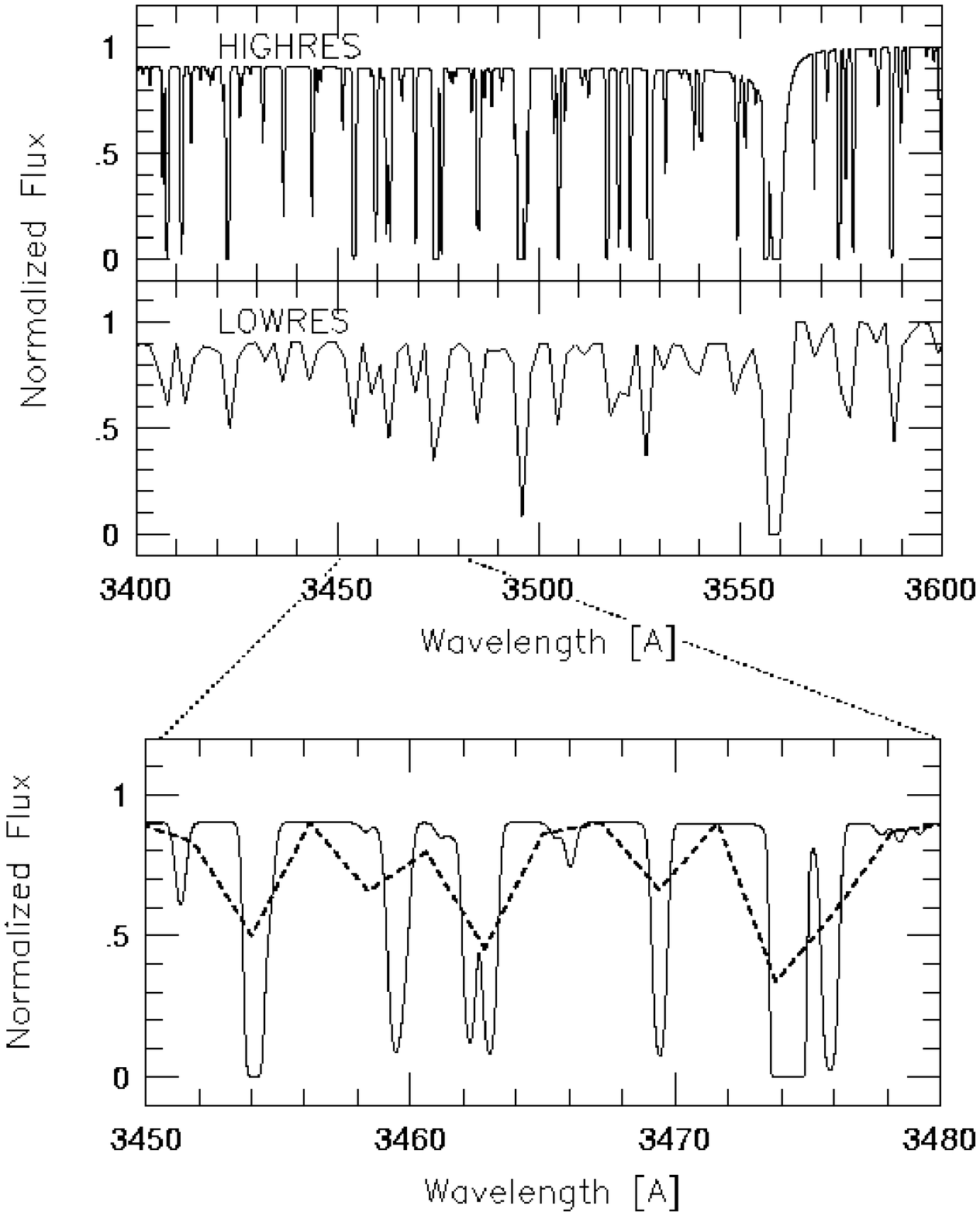}{8in}{0}{85}{85}{-250}{-25}
\caption{}
\end{figure}

\clearpage
\begin{figure}
\plotfiddle{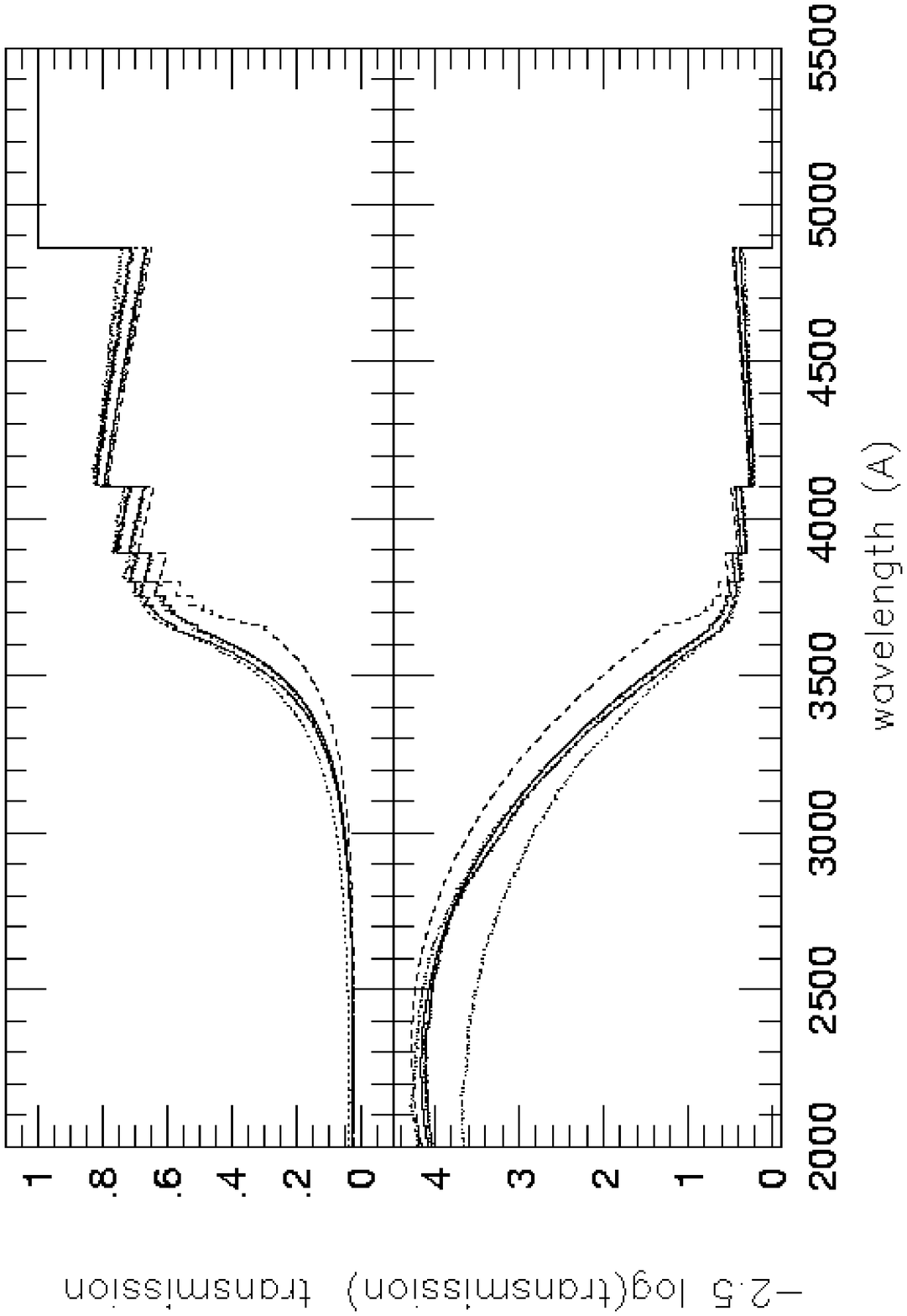}{8in}{0}{85}{85}{-250}{-25}
\caption{}
\end{figure}

\clearpage
\begin{figure}
\plotfiddle{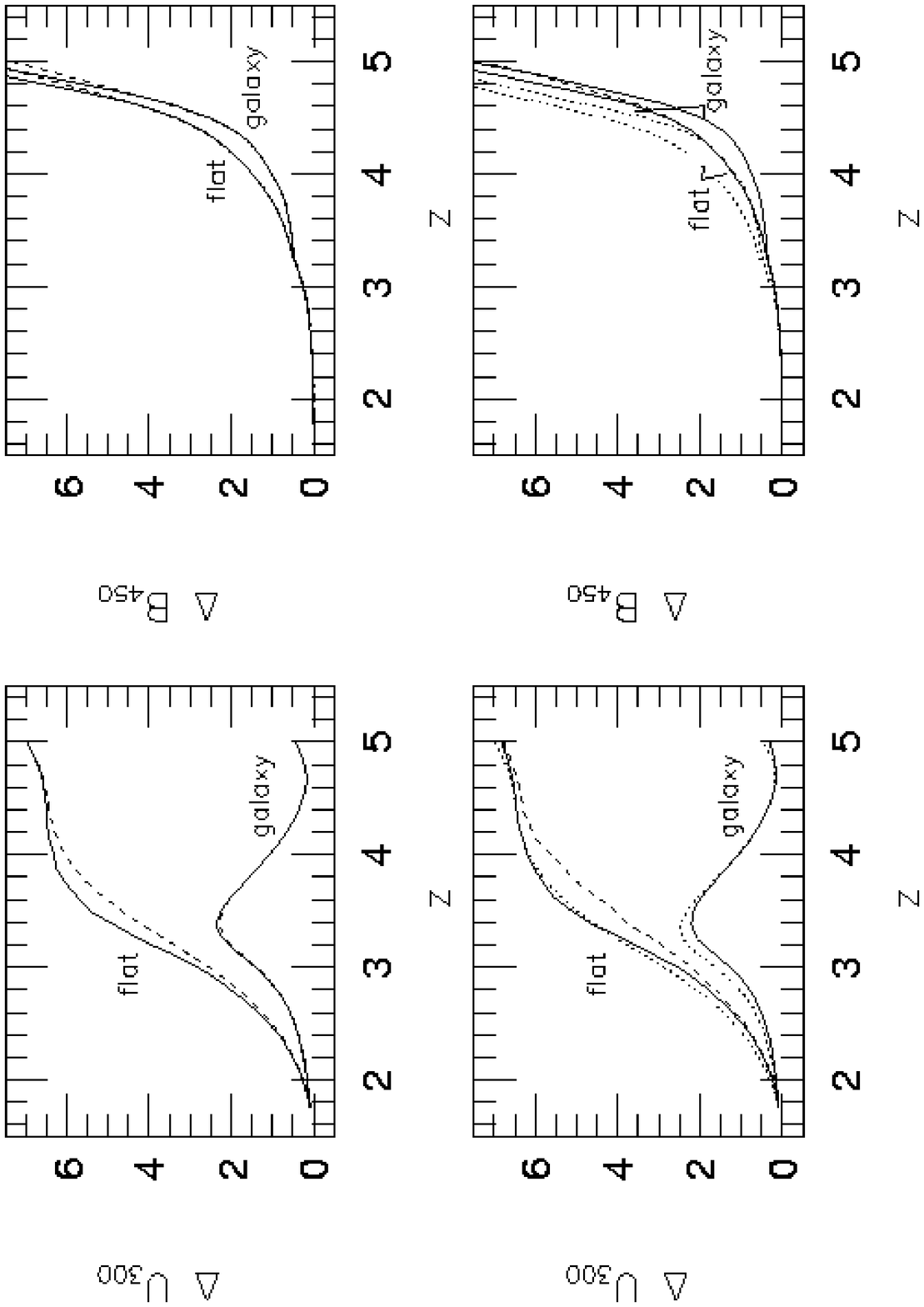}{8in}{0}{85}{85}{-250}{-25}
\caption{}
\end{figure}

\clearpage
\begin{figure}
\plotfiddle{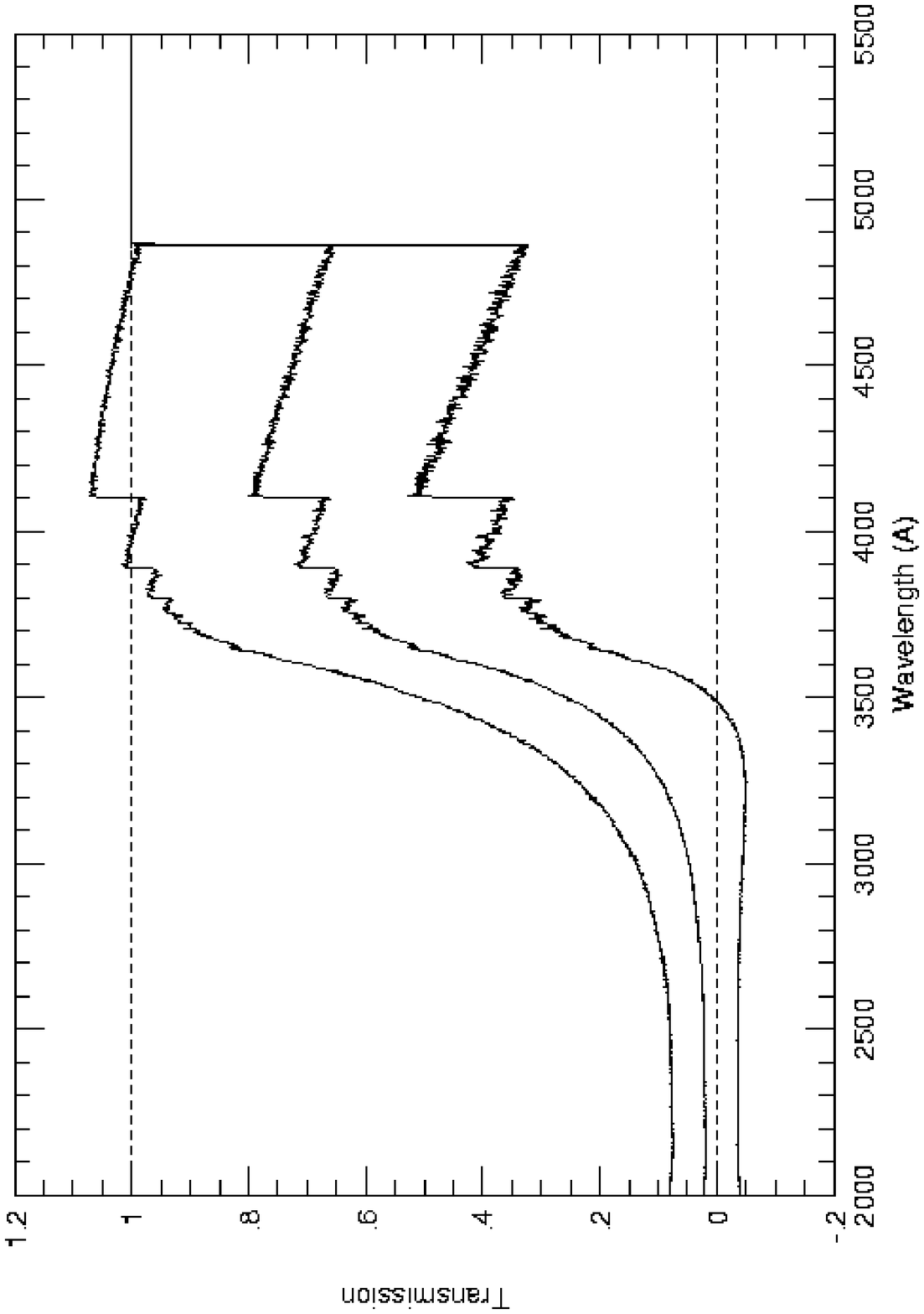}{8in}{0}{85}{85}{-250}{-25}
\caption{}
\end{figure}

\clearpage
\begin{figure}
\plotfiddle{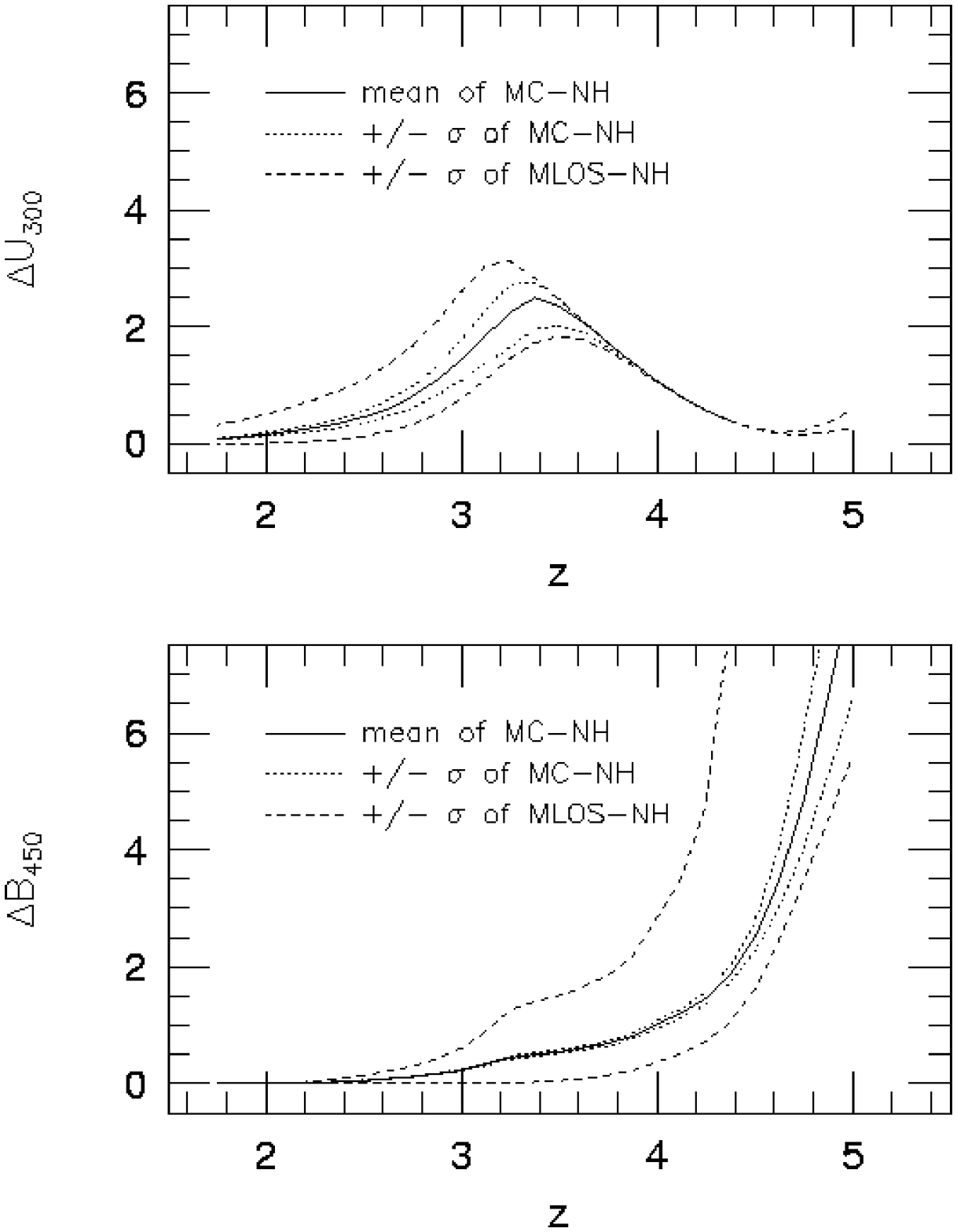}{8in}{0}{85}{85}{-250}{-25}
\caption{}
\end{figure}

\clearpage
\begin{figure}
\plotfiddle{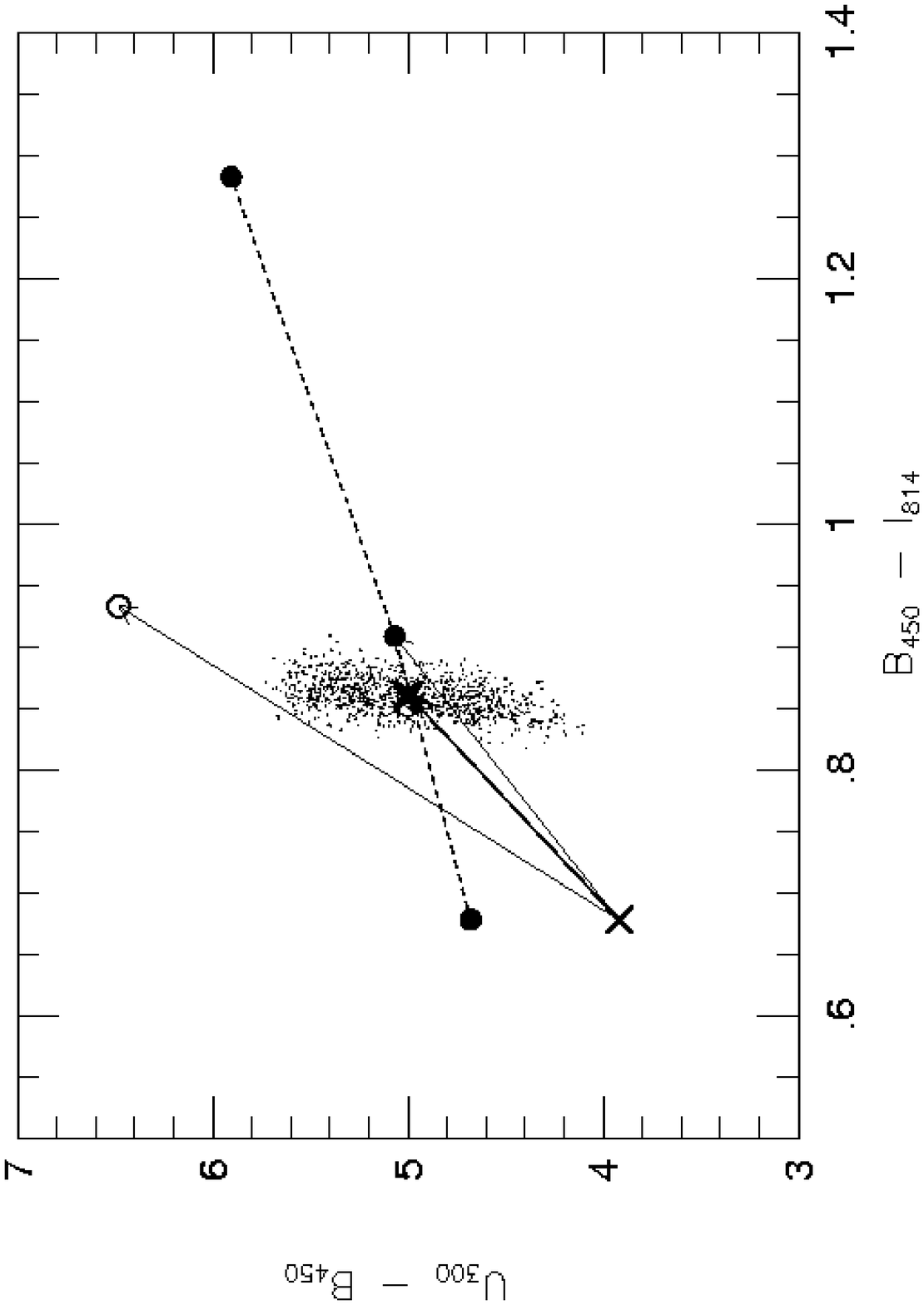}{8in}{0}{85}{85}{-250}{-25}
\caption{a}
\end{figure}

\setcounter{figure}{6} 

\clearpage
\begin{figure}
\plotfiddle{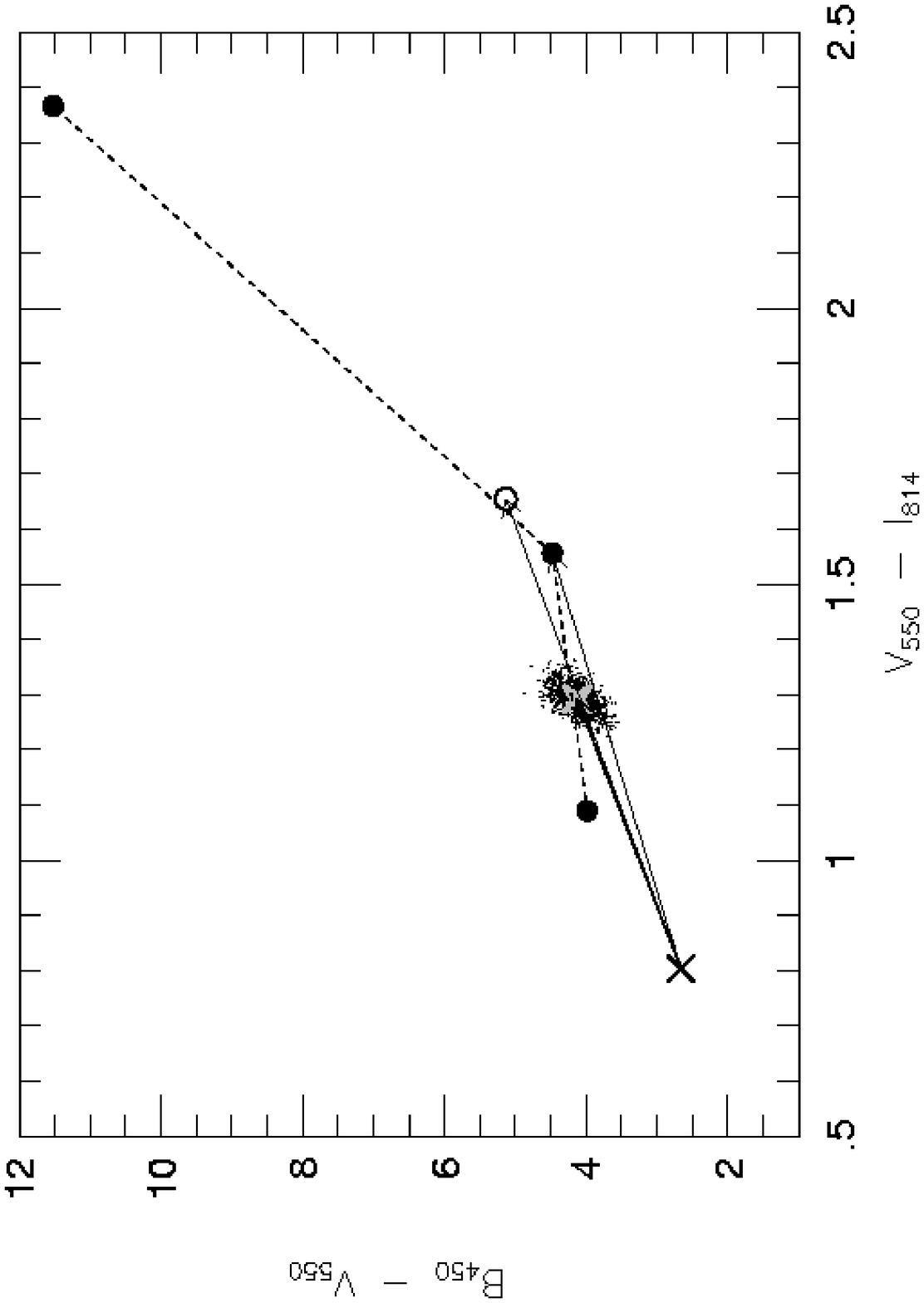}{8in}{0}{85}{85}{-250}{-25}
\caption{b}
\end{figure}

\clearpage
\begin{figure}
\plotfiddle{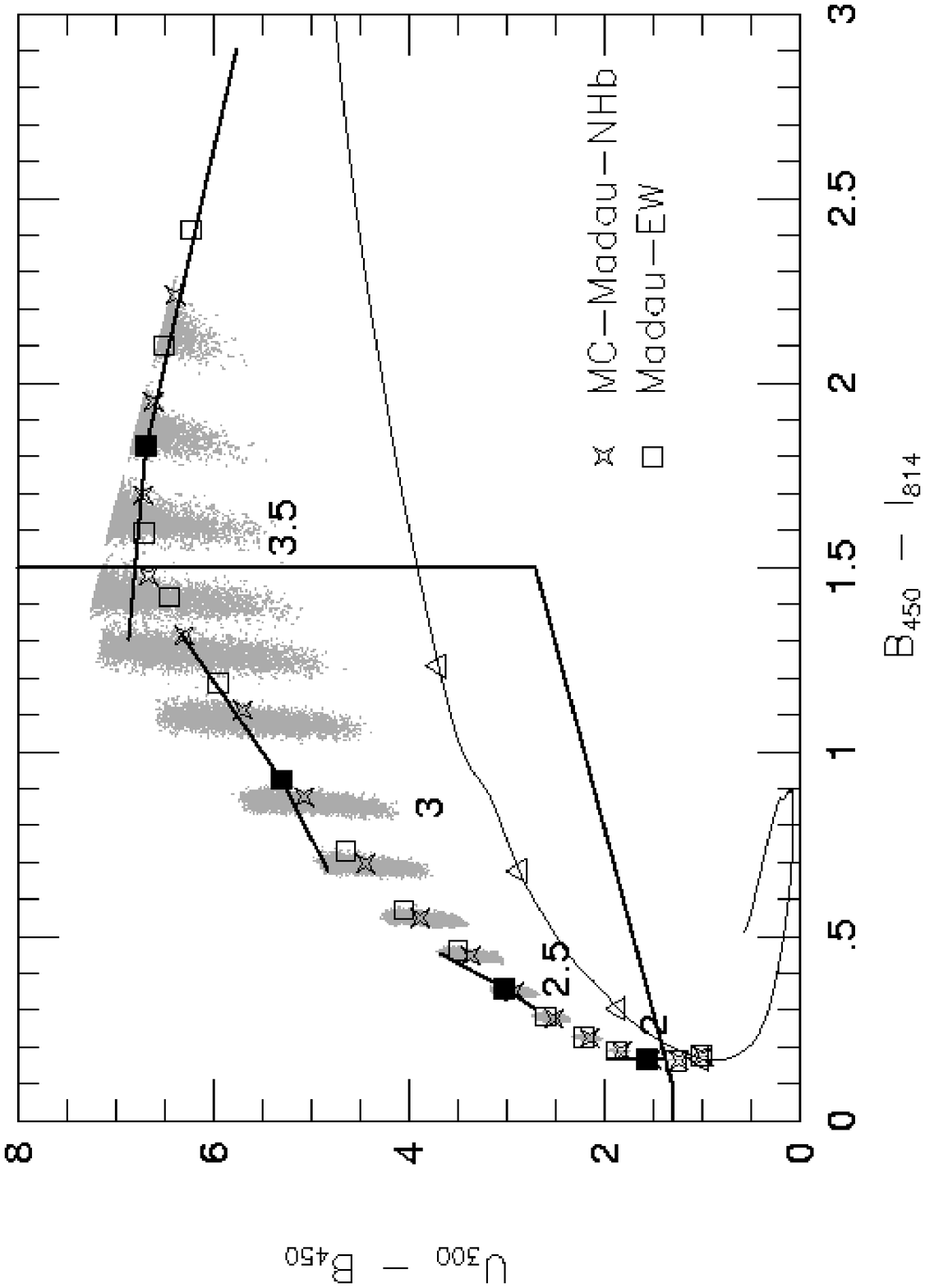}{8in}{0}{85}{85}{-250}{-25}
\caption{a}
\end{figure}

\setcounter{figure}{7} 

\clearpage
\begin{figure}
\plotfiddle{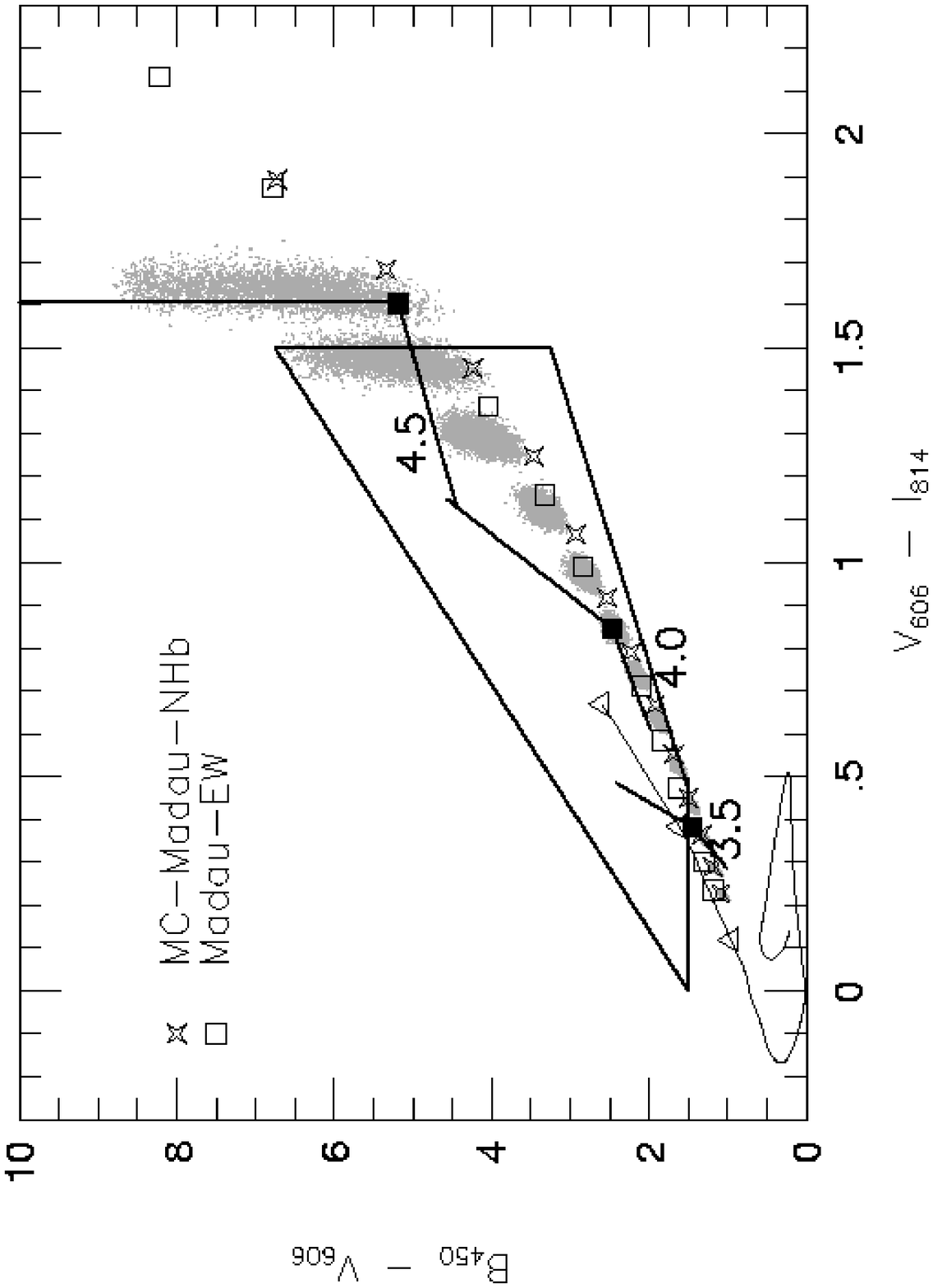}{8in}{0}{85}{85}{-250}{-25}
\caption{b}
\end{figure}

\clearpage
\begin{figure}
\plotfiddle{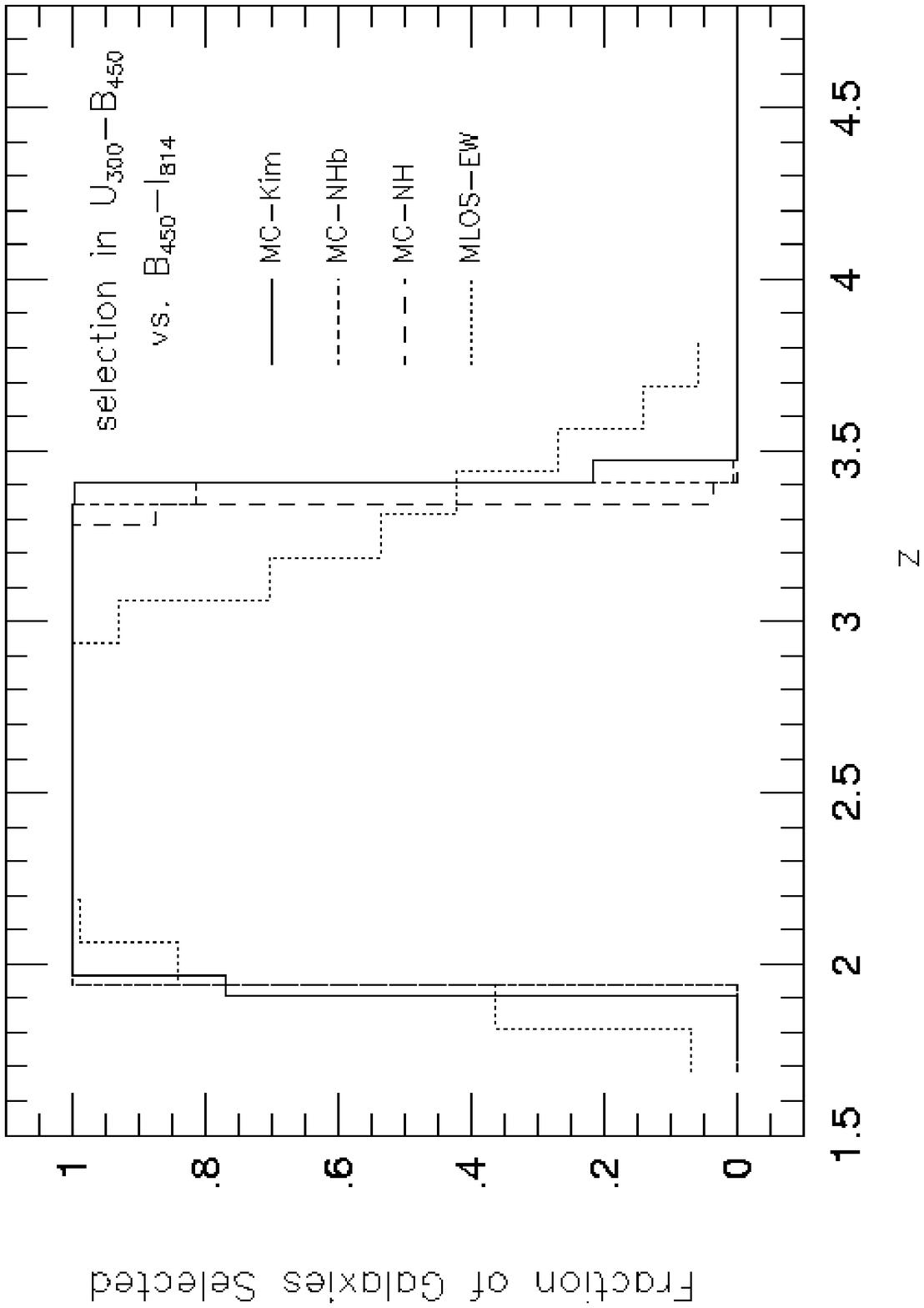}{8in}{0}{85}{85}{-250}{-25}
\caption{a}
\end{figure}

\setcounter{figure}{8} 

\clearpage
\begin{figure}
\plotfiddle{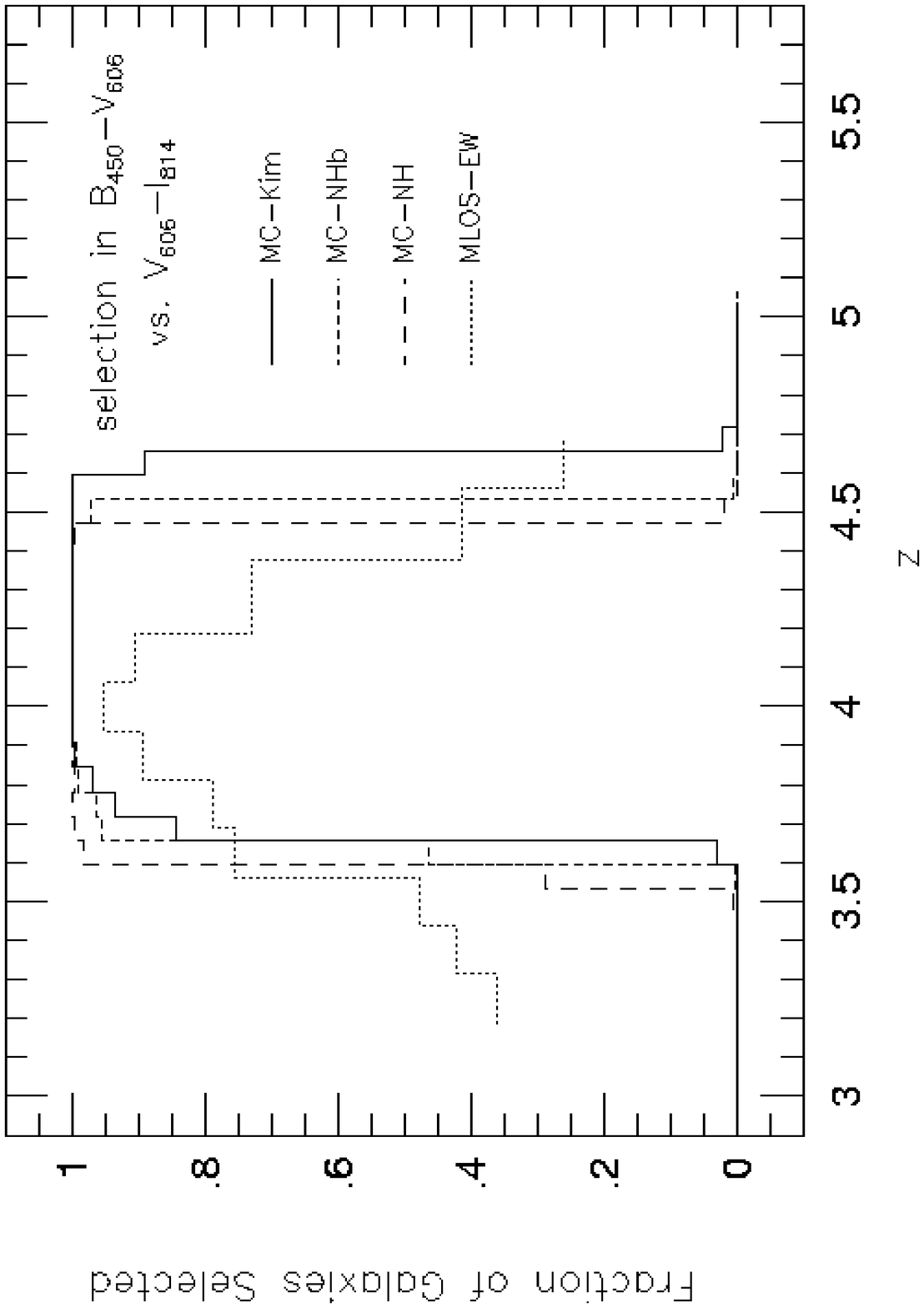}{8in}{0}{85}{85}{-250}{-25}
\caption{b}
\end{figure}

\clearpage
\begin{figure}
\plotfiddle{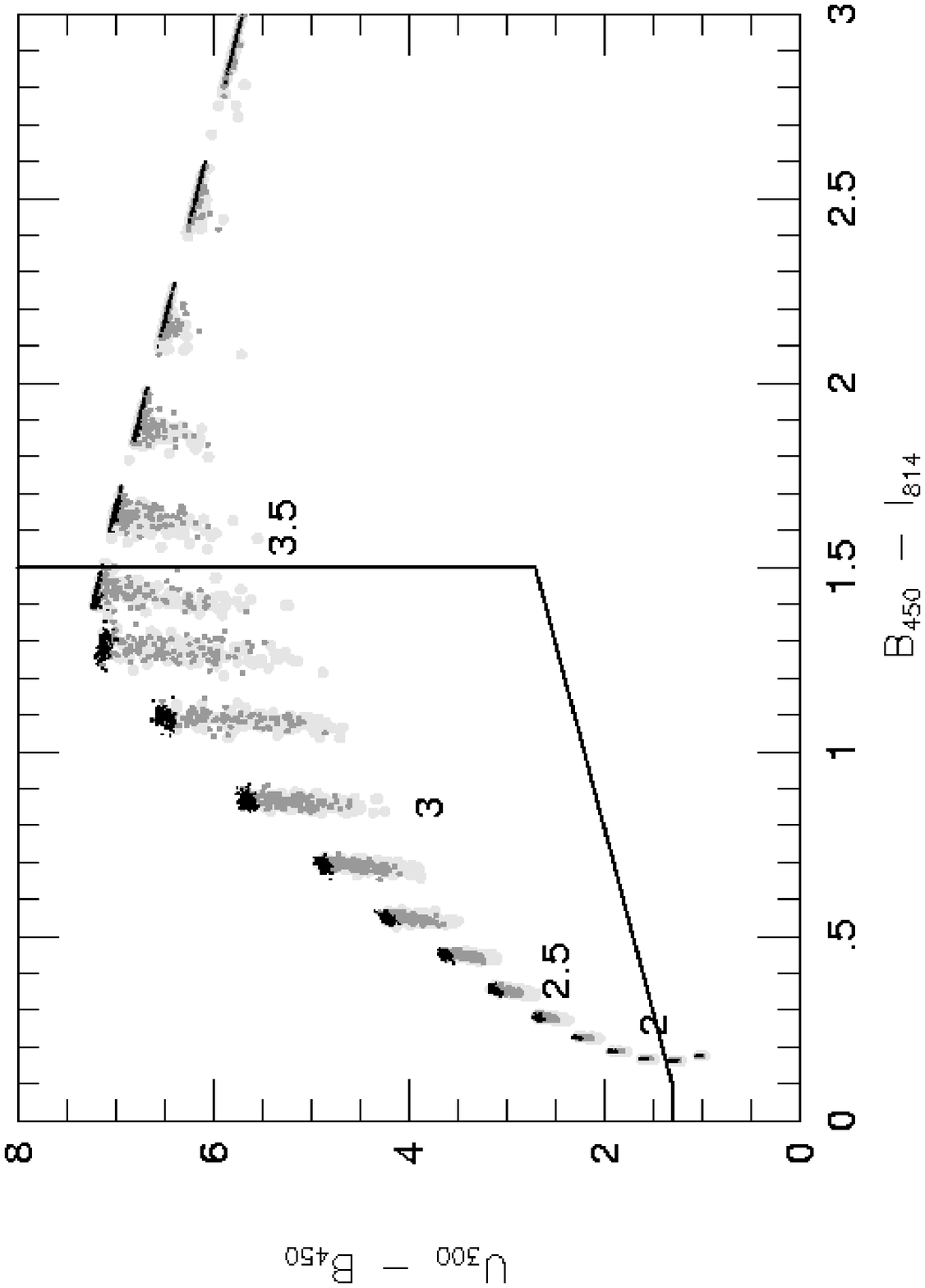}{8in}{0}{85}{85}{-250}{-25}
\caption{a}
\end{figure}

\setcounter{figure}{9} 

\clearpage
\begin{figure}
\plotfiddle{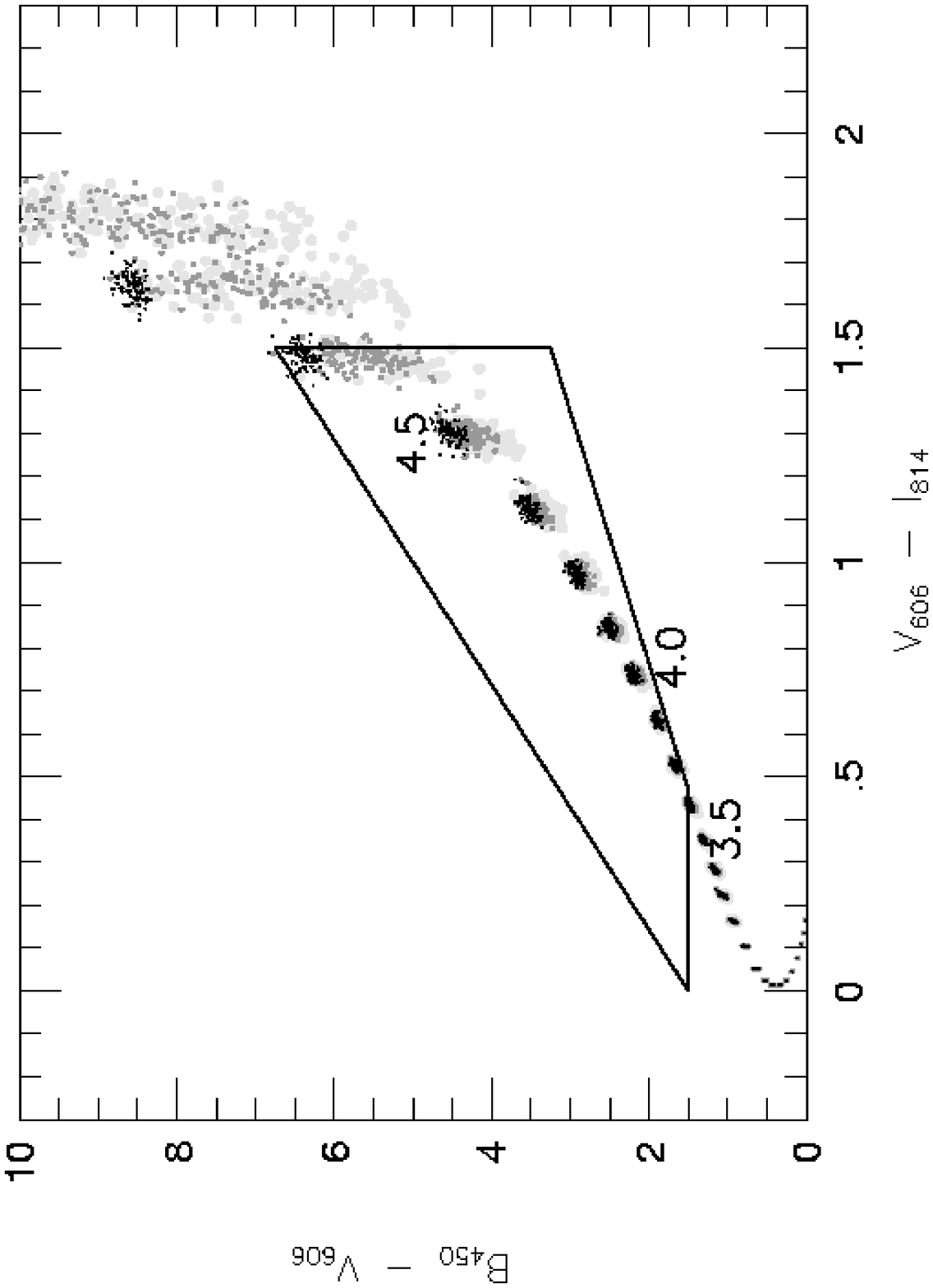}{8in}{0}{85}{85}{-250}{-25}
\caption{b}
\end{figure}


\begin{references}

\reference{} Bruzual, A. G. \& Charlot, S. 1993, ApJ, 405, 538

\reference{} Calzetti, D. 1997, 
in ``The Ultraviolet
Universe at Low and High Redshift'', ed. W. Waller, (Woodbury: AIP
Press), astro-ph/9706121

\reference{} Davis, M., Wilkinson, D. T. 1974, ApJ, 192, 251

\reference{} Deharveng, J.-M., Faisse, S., Milliard, B., Le Brun,
V. 1997, A\&A, 325, 1259

\reference{} Dickinson, M. 1998, to appear in ``The Hubble Deep
Field,'' eds. M. Livio, S.M. Fall, P. Madau, astro-ph/9802064

\reference{} Dove, J. B., Shull, J. M. 1994, ApJ, 423, 196

\reference{} Fall, S. M., Pei, Y. C. 1993, ApJ, 402, 479

\reference{} Ferguson, A. M. N., Wyse, R. F. G., Gallagher,
J. S. 1996, AJ, 112, 2567

\reference{} Giallongo, E., Trevese, D. 1990, ApJ, 353, 24


\reference{} Guhathakurta, P., Tyson, J. A., Majewski, S. R. 1990, ApJ, 357, L9

\reference{} Heisler, J., Ostriker, J. P. 1988, ApJ, 332, 543

\reference{} Kim, T.-S., Hu, E. M., Cowie, L. L., Songaila, A. 1997, AJ, 114, 1

\reference{} Koo, D. C. 1981, Ph.D. thesis, Berkeley 

\reference{} Koo, D. C. 1986, in ``The Spectral Evolution of
Galaxies,'' eds. C. Chiosi and A. Renzini (Dordrecht, Reidel), 419

\reference{} Koo, D. C., Kron, R. G. 1980, PASP, 92, 537

\reference{} Leitherer C., Ferguson, H. C., Heckman, T. M., Lowenthal,
J. D. 1995, ApJ, 454, L19

\reference{} Lockman, F. J., Jahoda, K., McCammon, D. 1986, ApJ, 302, 432

\reference{} Lowenthal, J. D. \etal 1997, ApJ, 481, 673

\reference{} Madau, P. 1995, ApJ, 441, 18

\reference{} Madau, P., Ferguson, H. C., Dickinson, M., Giavalisco, M.,
Steidel, C. C., Fruchter, A. 1996, MNRAS, 283, 1388

\reference{} Majewski, S. R. 1988, in ``Towards Understanding Galaxies
at Large Redshift,'' eds. R. G. Kron and A. Renzini, (Dordrecht,
Kluwer), 203

\reference{} Meier, D. L. 1976, ApJ, 207, 343

\reference{} Partridge, R. B. 1974, 192, 241

\reference{} Patel, K., Wilson C. D. 1995a, ApJ, 451, 607

\reference{} Patel, K., Wilson C. D. 1995b, ApJ, 453, 162

\reference{} Pettini, M. Steidel, C. C., Dickinson, M., Kellogg, M.,
Giavalisco, M., Adelberger, K. L.  1997, in ``The Ultraviolet
Universe at Low and High Redshift'', ed. W. Waller, (Woodbury: AIP
Press), astro-ph/9707200

\reference{} Reimers, D. \etal 1992, Nature, 360, 561

\reference{} Rudloff, K., and Baggett, S. 1995, anonymous ftp archive
${\rm http://www.stsci.edu/ftp/instrument\_news/WFPC2/Wfpc2\_thru}$


\reference{} Steidel, C.~C. \& Hamilton, D. 1993, AJ, 105, 2017

\reference{} Steidel, C. C., Pettini, M. \& Hamilton, D. 1995, AJ,
110, 2519 (S95)

\reference{} Steidel, C. C., Giavalisco, M., Pettini, M., Dickinson, M., \&
Adelberger, K. L. 1996, ApJ, 462, L17 (S96)

\reference{} Steidel, C. C., Adelberger, K. L., Dickinson, M.,
Giavalisco, M., Pettini, M., Kellogg, M. 1998, ApJ, 293, 428

\reference{} Vogt, S. S. \etal, 1994, SPIE, 2198, 362

\end{references}
\end{document}